\documentclass[journal]{IEEEtran}

\usepackage{amsmath,amssymb,amsfonts,amsthm,mathrsfs}
\usepackage{algorithm,algorithmic}
\usepackage{graphicx}
\usepackage{textcomp}
\usepackage{color,xcolor}
\usepackage{bm}
\usepackage{multirow}
\usepackage[normalem]{ulem}
\usepackage{tabularx}
\usepackage{xpatch}
\usepackage{soul}
\usepackage{ulem}
\usepackage{verbatim}
\usepackage{subcaption}
\usepackage{enumitem}
\usepackage{hyperref}
\usepackage[acronym]{glossaries}
\usepackage[linewidth=0.8pt]{mdframed}
\usepackage{cite}

\usepackage{array,booktabs}

\hypersetup{
	colorlinks=true,  
	linkcolor=red,    
	citecolor=blue,   
	urlcolor=black   
}

\usepackage{geometry}
\theoremstyle{theorem,lemma,remark,proposition}

\newtheorem{remark}{Remark}

\ifCLASSINFOpdf
\else
\fi

\begin{document}

\title{
Wideband Quantum Transduction for Rydberg Atomic Receivers Using Six-Wave Mixing
	
}

\author{Yuanbin~Chen,~Chau~Yuen,~\IEEEmembership{Fellow,~IEEE}~Chong~Meng~Samson~See,~\IEEEmembership{Member,~IEEE}
	

\thanks{Yuanbin Chen and Chau Yuen are with the School of Electrical and Electronics Engineering, Nanyang Technological University, Singapore 639798 (emails: yuanbin.chen@ntu.edu.sg; chau.yuen@ntu.edu.sg).

Chong Meng Samson See is with DSO National Laboratories, Singapore 118225 (e-mail: schongme@dso.org.sg).


}
	
	
}

\maketitle

\begin{abstract}
Rydberg atomic receivers hold extremely high sensitivity to electric fields, yet their effective 3-dB baseband bandwidth under conventional electromagnetically induced transparency (EIT) is typically constrained to tens to a few hundreds of kilohertz, which hinders wideband wireless applications. To relax this bottleneck, we investigate a six-wave mixing (SWM)-based Rydberg atomic receiver as a wideband radio frequency (RF)-to-optical quantum transducer. Specifically, we develop an explicit baseband input–output model that bridges the RF-induced atomic coherence to the detected optical readout.  
Based on the exact detected SWM response, we develop a reduced-order closed-form two-pole low-pass approximation under the near-resonant weak-signal of interest, which provides an analytical insight into how the 3-dB bandwidth is manipulated by the dressed higher-level atomic dynamics and optical/RF parameters. The validity range of this approximation is then quantified to clarify the operating conditions under which this reduced-order model accurately represents the exact SWM response.
We further characterize the linear dynamic range by employing the 1-dB compression point (P1dB) and the input-referred third-order intercept point (IIP3), unveiling a communication-compatible characterization of the bandwidth-sensitivity-linearity trade-off. Extensive simulation results demonstrate that SWM can achieve a 3-dB bandwidth of approximately 10~MHz while maintaining favorable linearity and sensitivity under the strict low-pass condition. The comparison with the EIT regime indicates that the two schemes should be treated as complementary rather than universally ordered. From an engineering perspective, the preferred SWM operating region is therefore not the one with the largest bandwidth, but the one that simultaneously provides a large bandwidth, acceptable sensitivity, favorable linearity, and low-pass regularity.

	
\end{abstract}


%
\IEEEpeerreviewmaketitle

\section{Introduction}
Quantum information technologies have matured from largely theoretical constructs into viable platforms for computing and secure communications, and another pillar has now come to the fore: quantum sensing~\cite{QS-101,QS-31}. By coherently mapping external stimuli, such as magnetic and electric fields, inertial forces, or tiny variations in gravity, onto well-controlled quantum degrees of freedom, quantum sensors can detect weak signals with sensitivities beyond the reach of classical systems. Among the most profound shifts driven by quantum innovation in wireless communications and sensing is the emergence of Rydberg atomic receivers~\cite{QS-103,QS-22}. These devices leverage the extreme sensitivity of highly excited Rydberg atoms, which function as radio frequency (RF)-to-optical quantum ``antennas" that transduce incident RF signals into optical signatures with unprecedented precision~\cite{terry-WCM,cui-CM,chen-arxiv,QS-1,QS-67}.

Rydberg atomic receivers depart fundamentally from conventional dipole-antenna-based RF front-ends. Instead of converting RF signals into currents in metal structures followed by low-noise amplifiers (LNA), a vapor of alkali atoms (typically rubidium or cesium) is optically driven into Rydberg states having enormous electric dipole moments and polarizabilities, so that even a minute RF field produces a resolvable Stark shift or Autler-Townes (AT) splitting in an electromagnetically induced transparency (EIT) resonance~\cite{QS-22,chen-arxiv2}. In other words, the atomic ensemble facilitates a coherent RF-to-optical conversion process: an incoming RF field disturbs the Rydberg manifold, and this disturbance is encoded onto the probe transmission spectrum as a readily detectable optical signal.
The optical readout of the EIT spectra allows simultaneous extraction of key parameters: the amplitude of the incident field~\cite{QS-12,QS-15,QS-20,QS-9-TAP-2024}, its polarization state~\cite{QS-24,QS-26,QS-54,QS-51}, and both the magnitude and orientation of an applied magnetic field~\cite{QS-52,QS-53}.
By leveraging EIT technique, the Rydberg atomic receivers having a single vapor cell have demonstrated the projected electric-field noise floor on the order of $\sim 0.01~\text{nV}/\text{cm}/\sqrt{\text{Hz}}$ in the standard quantum limit (SQL) regime~\cite{QS-34}, which is orders of magnitude far below the thermal noise floor of approximately $\sim 0.98~\text{nV}/\text{cm}/\sqrt{\text{Hz}}$~\cite{QS-39} used to benchmark classical receivers. Additionally, the dense manifold of Rydberg transitions supports carrier frequencies from near-DC up to the Terahertz (THz) regime within a single physical platform~\cite{QS-22}.

Despite these benefits, a significant constraint of current Rydberg atomic receivers is their intrinsically limited instantaneous bandwidth. In the canonical EIT-based scheme, the RF field slightly shifts the EIT resonance, so that its modulation is converted into small changes in the probe transmission and phase around the steep dispersion slope of a narrow transparency window~\cite{QS-16,QS-27,QS-28,QS-15,QS-18,QS-19,QS-20,QS-24,huizhi_wang,jian_xiao}.
The effective 3-dB RF bandwidth is therefore tied to the full-width at half-maximum (FWHM) of this EIT resonance, determined by the decoherence rates of the ground-state and Rydberg coherences, in addition to laser linewidth and transit-time broadening. To maintain quantum-limited sensitivity, one typically employs low probe power and moderate coupling strengths, which suppresses power broadening but additionally narrows the EIT window. Consequently, EIT-based Rydberg receivers often demonstrate practical baseband bandwidths limited to tens or a few hundreds of kilohertz, resulting in a fundamental tension between high sensitivity and the extensive signal bandwidths required for wideband wireless communications.

A viable approach to relax this bandwidth bottleneck is to move beyond simple three-level or four-level EIT and exploit higher-order nonlinear processes such as six-wave mixing (SWM) in multi-level Rydberg manifolds~\cite{SWM-2,SWM-2.5,QS-30,QS-30.5,ref_1_yang2024,ref_2_PRA}. Some representative SWM-based Rydberg studies have laid a solid physical foundation for microwave-to-optical conversion and wideband receiver design. Specifically, Han et al. demonstrate coherent microwave-to-optical conversion via SWM technique in cold $^{87}$Rb atoms, showing that the phase information of a microwave signal can be transferred to a generated optical field, with a conversion bandwidth exceeding 4~MHz~\cite{SWM-2}. Vogt et al. further improve the conversion efficiency to about $5\%$ in the linear regime~\cite{SWM-2.5}.
More recent wideband-oriented studies have shown that SWM can substantially broaden the receiver response: Yang et al. report a Rydberg superheterodyne receiver with a sensitivity of $\sim 62~\text{nV}/\text{cm}/\sqrt{\text{Hz}}$ and an instantaneous bandwidth up to $\pm 10.2~\text{MHz}$, attributed to the amplification of one generated sideband wave in the SWM process~\cite{ref_1_yang2024}; Shylla et al. experimentally resolve the positive and negative sidebands arising from distinct SWM pathways and show that the negative sideband can reach a larger 3-dB bandwidth of about $11~\text{MHz}$ under optimized parameters~\cite{ref_2_PRA}; and Bor\'owka et al. demonstrate a continuous room-temperature Rydberg microwave-to-optical converter based on free-space SWM, achieving a conversion bandwidth of $16~\text{MHz}$~\cite{QS-30}.

Despite these impressive advances, the existing SWM literature primarily establishes physical feasibility, conversion efficiency, sideband generation, or experimentally observed bandwidth. What remains missing, from an engineering receiver-design perspective, is a quantitative framework that determines when such bandwidth can be actually treated as a detected 3-dB bandwidth. 
This distinction is crucial for practical implementations of Rydberg atomic receivers. More importantly, a larger spectral or conversion bandwidth alone does not guarantee a useful wideband front-end if it is accompanied by degraded responsivity, harsh noise-equivalent field, gain compression, or inter-modulation distortion. Therefore, practical implementation requires not only demonstrating SWM-based bandwidth enhancement, but also establishing a receiver-level design criterion that jointly evaluates bandwidth, sensitivity, linearity, and response regularity. To the best of our knowledge, this work provides the first such communication-oriented operating-region characterization for SWM-based Rydberg atomic receivers. Our main contributions are summarized as follows.

\begin{itemize}
\item We formulate a six-level SWM-based Rydberg atomic receiver as a wideband RF-to-optical quantum transducer, in which the probe, coupling, local oscillator (LO), and auxiliary optical fields, together with the RF signal, drive a closed six-wave-mixing loop that generates an output optical field whose complex envelope carries the full RF information. Beginning with the Hamiltonian and master equation, we eliminate intermediate coherences to derive a closed-form expression for the SWM coherence and fifth-order polarization. Next, we develop an explicit wideband baseband model spanning from the probe input to the output light field. This yields, to the best of our knowledge, the first compact input–output baseband model for SWM-based Rydberg atomic receivers, bridging high-order nonlinear atomic dynamics with a communication-oriented system for wideband quantum transduction analysis.

\item Relying upon this model, under the near-resonant weak-signal regime of interest, the exact detected SWM response can be reduced to a closed-form two-pole low-pass approximation. This reduced-order model clarifies how the detected 3-dB bandwidth is determined by the dressed higher-level atomic dynamics and the relevant optical/RF parameters. To justify the employment of this approximation, we further examine its validity range through approximation-error and low-pass-regularity analyses. Additionally, to quantify the linear dynamic range of the SWM-based Rydberg atomic receiver considered, we analyze the 1-dB compression point (P1dB) and the input-referred third-order intercept point (IIP3), standard RF figures of merit that mark the onset of gain compression and third-order inter-modulation distortion in the RF-to-optical transduction chain.

\item We numerically evaluate the proposed SWM-based Rydberg atomic receiver using the QuTiP toolbox~\cite{QuTip} and compare it with a conventional four-level EIT-based configuration. We first evince that the proposed reduced-order approximation accurately captures both the detected bandwidth and the passband shape within properly selected operating regions. Excessive detuning or overly strong dressing may induce undesirable non-low-pass features and consequently degrade the approximation accuracy. Then, our simulation 
results demonstrate that SWM can achieve a 3-dB bandwidth of approximately 10~MHz, compared with the sub-MHz bandwidth obtained with the EIT regime, while maintaining favorable linearity and sensitivity under the strict low-pass condition.
The auxiliary field in the SWM configuration serves as an effective bandwidth-control knob, but its strength must be carefully selected under the low-pass regularity constraint to avoid distortion effect.

\item From an engineering perspective, the comparison with the EIT regime indicates that this pair of schemes should be treated as complementary rather than subject to a universal ranking. EIT remains attractive for narrowband sensing and reception due to its stronger local responsivity. In contrast, the key benefit of SWM lies in that it expands the feasible design space: relying upon the proposed closed-form two-pole low-pass framework, SWM enables wideband operation while retaining high sensitivity and favorable linearity, provided that the auxiliary-field strength is appropriately selected.
Importantly, the preferred SWM operating region is not the one that maximizes bandwidth alone, but rather the one that jointly provides a large bandwidth, acceptable noise-equivalent field (NEF), favorable IIP3, and low-pass regularity.


\end{itemize}


The remainder of this paper is structured as follows. Sec.~\ref{Sec_II_system_model} introduces the fundamentals of the six-level SWM-based Rydberg atomic receiver and derives the RF-to-optical baseband transfer response. Sec.~\ref{Sec_III} analyzes the small-signal response and linearity, yielding closed-form expressions for the 3-dB bandwidth, P1dB, and IIP3 of the proposed architecture. Sec.~\ref{Sec_IV_simulation} presents numerical results to validate our analytical model, to compare the SWM and EIT schemes, and to elucidate the bandwidth-linearity trade-offs. Finally, Sec.~\ref{Sec_V_conclusion} concludes the paper.

\section{System Model}\label{Sec_II_system_model}

\begin{figure*}[t]
	\centering
	\includegraphics[width=\textwidth]{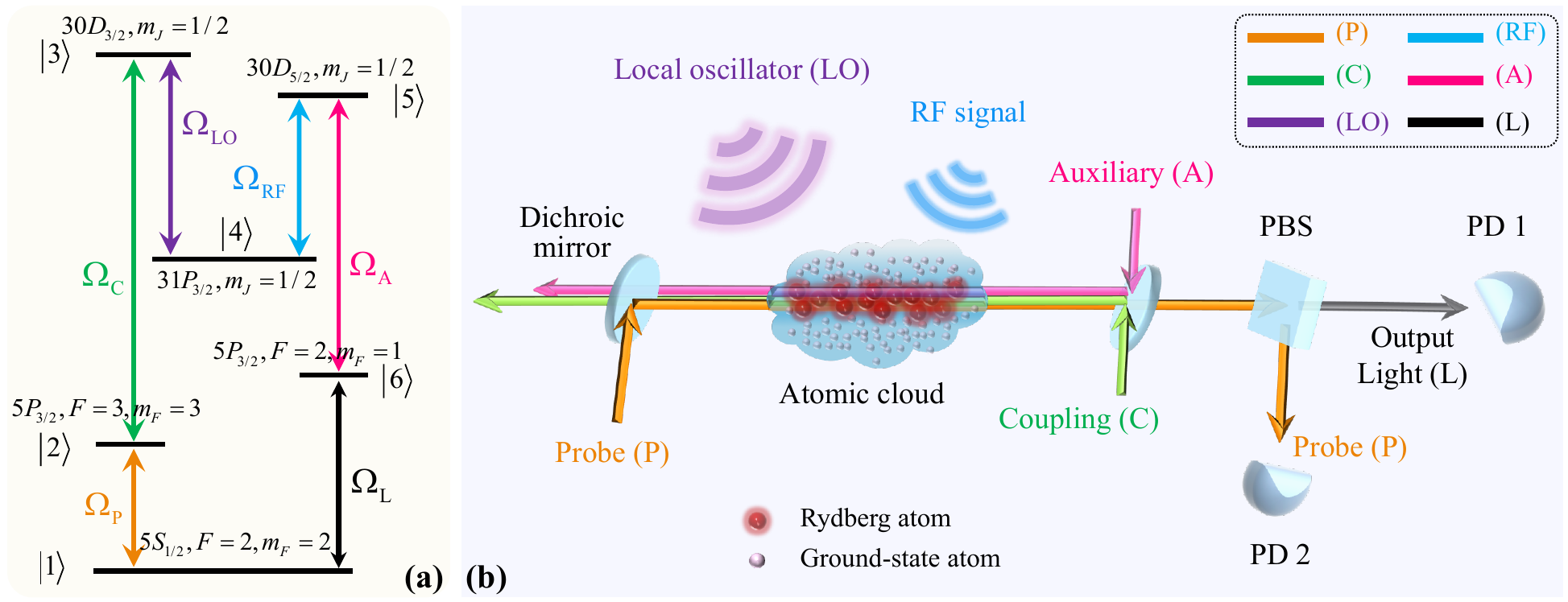}
	\caption{Illustration of the SWM-based Rydberg atomic receiver. (a) Six-level energy transition diagram. (b) The RF signal is converted to the (output) light field  by SWM-based Rydberg atomic receiver.} \label{SWM_model}
\end{figure*}

\subsection{Six-Level Atomic Systems and SWM Architecture}
The energy levels for the SWM are shown in Fig.~\ref{SWM_model}(a), and the SWM-based Rydberg atomic receiver is sketched in Fig.~\ref{SWM_model}(b). In our configuration, the conversion of the input RF signal into the output light field~(L) is achieved via frequency mixing with four driving fields: the probe field~(P), the coupling field~(C), the strong LO, and the auxiliary optical field (A), in a cloud of cold $^{87}\text{Rb}$ atoms. In this regime, the atomic thermal velocity is sufficiently small so that Doppler broadening and velocity averaging can be neglected, and collision/pressure broadening is also strongly suppressed compared with hot-vapor cells.
Starting from the ground state $\left| 1 \right\rangle $, these optical fields together with the RF signal, all tuned close to their respective atomic transitions, build up a coherence between the initial state $\left| 1 \right\rangle $ and the highest Rydberg state $\left| 6 \right\rangle $. Under a weak time-varying RF perturbation, the atomic coherence induced by the closed SWM loop acquires a corresponding modulation component, so that the RF signal envelope is imprinted onto the atomic superposition and, thus, onto the output light field (L). This modulated optical field is not the final observable by itself; rather, it is subsequently converted into a detected heterodyne current through interference with the probe field. In this way, the SWM process serves as a coherent transducer that maps small RF-induced changes of the Rydberg coherence into measurable optical signatures.
Furthermore, the P and L fields that emerge from the atomic cloud are collected by a separated employing a polarization beam splitter (PBS), as portrayed in Fig.~\ref{SWM_model}(b). Their respective powers can be measured with two different avalanche photodiode
detectors~(PDs), i.e., PD~1 and PD~2 illustrated in Fig.~\ref{SWM_model}(b).
The output light field (L) has a frequency of
\begin{equation}
{\omega _{\text{L}}} = {\omega _{\text{P}}} + {\omega _{\text{C}}} - {\omega _{\text{A}}} + {\omega _{{\text{LO}}}} - {\omega _{{\text{RF}}}},
\end{equation}
which ensures that the resonant six-wave-mixing loop
$\left| 1 \right\rangle \mathop  \to \limits^{{\Omega _{\text{P}}}} \left| 2 \right\rangle \mathop  \to \limits^{{\Omega _{\text{C}}}} \left| 3 \right\rangle \mathop  \to \limits^{{\Omega _{{\text{LO}}}}} \left| 4 \right\rangle \mathop  \to \limits^{{\Omega _{{\text{RF}}}}(t)} \left| 5 \right\rangle \mathop  \to \limits^{\Omega _{\text{A}}^*} \left| 6 \right\rangle \mathop  \to \limits^{{\Omega _{\text{L}}}} \left| 1 \right\rangle $
is closed. Here, $\omega_\text{X}$ and $\Omega_\text{X}$ denote the angular frequency and Rabi frequency of field ${\text{X}} \in \left\{ {{\text{P}},{\text{C}},{\text{A}},{\text{LO}},{\text{RF}},{\text{L}}} \right\}$, respectively.

The output optical field (L) is determined by the corresponding phase-matching condition given by
\begin{equation}
{{\mathbf{k}}_{\text{L}}} = {{\mathbf{k}}_{\text{P}}} + {{\mathbf{k}}_{\text{C}}}  - {{\mathbf{k}}_{\text{A}}} + {{\mathbf{k}}_{{\text{LO}}}} - {{\mathbf{k}}_{{\text{RF}}}} ,
\end{equation}
where ${\mathbf{k}}_{\text{X}}$ is the wave vector of field X.

\subsection{Hamiltonian and Master Equation}

Under the rotating-wave approximation, the coherent evolution is characterized by a time-dependent Hamiltonian ${\mathbf{\hat H}}$ acting on the basis $\left\{ {\left| 1 \right\rangle ,...,\left| 6 \right\rangle } \right\}$. The Hamiltonian can be written in a compact form as
\begin{equation}\label{Hamiltonian}
	{\mathbf{\hat H}} =  - \frac{\hbar }{2}
	\left( {\begin{array}{*{20}{c}}
			{\;0}&{{\Omega _\text{P}}}&0&0&0&0 \\ 
			{\Omega _\text{P}^{*}}&{2{\Delta _2}}&{{\Omega _\text{C}}}&0&0&0 \\ 
			0&{\Omega _\text{C}^{*}}&{2{\Delta _3}}&{{\Omega _\text{LO}}}&0&0 \\ 
			0&0&{\Omega _\text{LO}^{*}}&{2{\Delta _4}}&{{\Omega _\text{RF}}\left( t \right)}&0 \\ 
			0&0&0&{\Omega _\text{RF}^{*}\left( t \right)}&{2{\Delta _5}}&{{\Omega _\text{A}}} \\ 
			0&0&0&0&{\Omega _\text{A}^{*}}&{2{\Delta _6}} 
	\end{array}} \right).
\end{equation}
where $\Delta_\text{X}$ represents the single-photon detuning of field ${\text{X}} \in \left\{ {{\text{P}},{\text{C}},{\text{A}},{\text{LO}},{\text{RF}},{\text{L}}} \right\}$, while detunings associated with levels $\left| 2 \right\rangle ,...,\left| 6 \right\rangle $ are defined as ${\Delta _2} = {\Delta _{\text{P}}}$, ${\Delta _3} = {\Delta _{\text{P}}} + {\Delta _{\text{C}}}$, ${\Delta _5} = {\Delta _{\text{P}}} + {\Delta _{\text{C}}} - {\Delta _{{\text{LO}}}} + {\Delta _{{\text{RF}}}}$, ${\Delta _4} = {\Delta _{\text{P}}} + {\Delta _{\text{C}}} - {\Delta _{\text{LO}}}$ and ${\Delta _6} = {\Delta _{\text{L}}} \approx 0$.

The atomic coherences are given by the steady-state solution $\bm{\rho}$ of the following Lindblad master equation (also known as the optical Bloch equation)~\cite{QS-15}
\begin{equation}\label{master_equation}
	{\partial _t} {\bm{\rho}} =  - \frac{\jmath }{\hbar }\left[ {{\mathbf{\hat H}},\bm{\rho}} \right] + {\mathcal{L}_\Gamma }\left( \bm{\rho } \right) + {\mathcal{L}_{{\text{depth}}}}\left( \bm{\rho } \right).
\end{equation}
The Lindblad term  ${\mathcal{L}_\Gamma }\left( \bm{\rho } \right)$ (also known as population-decay Lindblad dissipator) accounts for spontaneous emission from the intermediate states $\left| 2 \right\rangle $ and $\left| 6 \right\rangle $  and from the Rydberg states, which is given by
\begin{equation}
{\mathcal{L}_\Gamma }\left( \bm{\rho } \right) = \sum\limits_{j > i} {{\Gamma _{j \to i}}\mathcal{D}\left[ {\left| i \right\rangle \left\langle j \right|} \right] \bm{\rho }} ,
\end{equation}
\begin{equation}
\mathcal{D}\left[ {\left| i \right\rangle \left\langle j \right|} \right]\bm{\rho} = \left[ {\left| i \right\rangle \left\langle j \right|} \right]\bm{\rho }{\left[ {\left| i \right\rangle \left\langle j \right|} \right]^\dag } - \frac{1}{2}\left\{ {{{\left[ {\left| i \right\rangle \left\langle j \right|} \right]}^\dag }\left[ {\left| i \right\rangle \left\langle j \right|} \right],\bm{\rho }} \right\}.
\end{equation}
The term ${\mathcal{L}_{{\text{depth}}}}\left( \bm{\rho } \right)$ (also known as pure-dephasing Lindblad dissipator) accounts for additional dephasing mechanisms and can be written as
\begin{equation}
{\mathcal{L}_{{\text{depth}}}}\left( \bm{\rho } \right) = \sum\limits_{j > i} {\frac{1}{2}\left[ {{\gamma _{ij}} - \frac{1}{2}\left( {{\Gamma _i} + {\Gamma _j}} \right)} \right]\mathcal{D}\left[ {\left| i \right\rangle \left\langle i \right| - \left| j \right\rangle \left\langle j \right|} \right]\bm{\rho }} ,
\end{equation}
where ${\gamma _{ij}}$ is the coherence dephasing rate and  ${\Gamma _i}$ represents the population decay rate associated with state $\left| j \right\rangle$. These rates include the effects of atomic collisions, dipole-dipole interactions between Rydberg atoms, and finite laser linewidths. The dephasing rates $\gamma_1$, $\gamma_2$, and $\gamma_6$ are neglected since they are much smaller than $\Gamma$~\cite{SWM-2, SWM-2.5}.

To obtain the frequency response of the SWM, we solve the linearized master equation for the first-order density matrix component $\rho_{j1}$. The steady-state solution $\rho_{61}$ can be explicitly given by
\begin{align}\label{rho_61}
{\rho _{61}}\left( \omega  \right)  = {\left( {\frac{\jmath }{2}} \right)^5} \ & \frac{{{\Omega _{\text{P}}}{\Omega _{\text{C}}}{\Omega _{{\text{LO}}}}}}{{{D_2 \left( \omega  \right)}{D_3 \left( \omega  \right)}{D_4 \left( \omega  \right)}}} \nonumber\\
 \times &  \frac{{\Omega _A^*{\Omega _{{\text{RF}}}}\left( \omega  \right)}}{{{D_5}\left( \omega  \right){D_6}\left( \omega  \right) + \frac{{{{\left| {{\Omega _A}} \right|}^2}}}{4}}},
\end{align}
in which ${D_j}\left( \omega  \right) = D_j^{\left( 0 \right)} + \jmath \omega  $, $D_j^{\left( 0 \right)} \triangleq {\gamma _{j1}} + \jmath {\Delta _{j1}},j = 2,...,6$, represents the complex detuning, $\gamma_{j1}$ is the decoherence rate (including both population decay and pure dephasing) of $\rho_{j1}$, and $\Delta_j$ represents the detuning of level $\left| j \right\rangle $ from the multi-photon resonance. A detailed derivation of $\rho_{61}$ is provided in Appendix~\ref{appendix_rho_61}.
To make the role of the lower-level coherences explicit, we further rewrite
\begin{equation}\label{D2_D3_D4}
\frac{1}{{{D_2}\left( \omega  \right){D_3}\left( \omega  \right){D_4}\left( \omega  \right)}} = \frac{1}{{D_2^{\left( 0 \right)}D_3^{\left( 0 \right)}D_4^{\left( 0 \right)}}}\prod\limits_{j = 2}^4 {{{\left( {1 + \frac{{\jmath \omega }}{{D_j^{\left( 0 \right)}}}} \right)}^{ - 1}}} .
\end{equation}
This formulation shows that $\rho_{61}$ is not, in general, determined solely by ${D_5}\left( \omega  \right)$ and ${D_6}\left( \omega  \right)$. Instead, the lower-level coherences contribute an additional frequency-dependent factor. In the near-resonant weak-signal regime of interest, and over a sufficiently narrow modulation band around the operating point, this factor, i.e., $1 / \left[ {{D_2}\left( \omega  \right){D_3}\left( \omega  \right){D_4}\left( \omega  \right)} \right]$, may vary slowly compared with the final-stage dressed kernel~\cite{SWM-2.5,ref_1_yang2024}. In this sense, the dominant roll-off of the detected SWM response is determined by the ${D_5} - {D_6}$ subsystem, while the contribution is absorbed into an approximately constant complex gain. An error analysis for this approximation is detailed in Appendix~\ref{appendix_error_analysis}.


Regarding ${\rho _{61}}\left( \omega  \right)$, in a density-matrix language, this parameter describes how strongly the atoms oscillate between levels $\left| 6 \right\rangle$ and $\left| 1 \right\rangle$  at the analysis frequency $\omega$. Physically, this oscillation creates an electric dipole on the $\left| 6 \right\rangle  \to \left| 1 \right\rangle $ transition. A larger ${\rho _{61}}\left( \omega  \right)$ leads to a stronger dipole oscillation, which in turn enhances the radiated light on this transition. When we consider an ensemble atomic cloud, these dipoles add up to a macroscopic polarization at the generated-light frequency $\omega_{\text{L}}$. This polarization is the actual source term that appears in Maxwell's equations and drives the output light field~(L)~\cite{SWM-1}. Therefore, once ${\rho _{61}}\left( \omega  \right)$ is known, the next step is to convert it into the corresponding polarization and then into the light field~(L) at the output of the atomic cloud. This provides a direct link between the RF-induced atomic dynamics and the measurable optical signal, and ultimately defines the RF-to-baseband transfer function.

\subsection{Optical Readout and Baseband Model}
To quantitatively characterize the light field (L) at the output of the cold atomic cloud, we denote the atomic response by a fifth-order nonlinear polarization ${P^{\left( 5 \right)}}\left( \omega  \right)$, because the output field~(L) arises from mixing the minute RF field with four auxiliary fields (P, C, LO, and A). The fifth-order nonlinear polarization ${P^{\left( 5 \right)}}\left( \omega ,z \right)$ is proportional to the coherence $\rho_{61} \left( \omega \right) $, which has the form of
\begin{equation}
{P^{\left( 5 \right)}}\left( \omega,z  \right) = N \left( z \right) {\mu _{61}}{\rho _{61}}\left( \omega  \right),
\end{equation}
where $N\left(z \right) $ denotes the atomic number density along the propagation axis $z$, and ${\mu _{61}}$ is the dipole moment associated with transition $\left| 6 \right\rangle  \to \left| 1 \right\rangle $. 
Physically, ${P^{\left( 5 \right)}}\left( \omega,z  \right)$  represents the effective ``RF-driven dipole strength" of the whole atomic dynamics at position $z$, that is to say, it directly determines how efficiently the incident RF signal is up-converted into the output light field (L).
By substituting $\rho_{61} \left(\omega \right) $ in (\ref{rho_61}), we obtain 
\begin{align}\label{P_5}
&{P^{\left( 5 \right)}}\left( \omega,z  \right) \nonumber\\
& = N \left( z \right) {\mu _{61}}{\left( {\frac{\jmath }{2}} \right)^5} \ \frac{{{\Omega _{\text{P}}}{\Omega _{\text{C}}}{\Omega _{{\text{LO}}}}}}{{{D_2}{D_3}{D_4}}}\frac{{\Omega _{\text{A}}^*{\Omega _{{\text{RF}}}}\left( \omega  \right)}}{{{D_5}\left( \omega  \right){D_6}\left( \omega  \right) + \frac{{{{\left| {{\Omega _{\text{A}}}} \right|}^2}}}{4}}},
\end{align}
where we have made explicit the dependence on the local density $N \left( z \right)$. Equivalently, by leveraging the usual relation between Rabi frequencies and electric fields, i.e., ${\Omega _{\text{X}}} = {\mu _{\text{X}}}{E_{\text{X}}}/\hbar $ ($E_{\text{X}}$ for the electric-field amplitude and $\hbar$ for reduced Planck constant) for each driven transition, ${P^{\left( 5 \right)}}\left( \omega,z  \right)$ can be restructured as 
\begin{align}
&{P^{\left( 5 \right)}}\left( {\omega ,z} \right) \nonumber\\
&= {\epsilon_0}\chi _{{\text{eff}}}^{\left( 5 \right)}\left( {\omega ,z} \right){E_{\text{P}}}\left( z \right){E_{\text{C}}}\left( z \right){E_{{\text{LO}}}}\left( z \right)E_{\text{A}}^*\left( z \right){E_{{\text{RF}}}}\left( \omega  \right),
\end{align}
where $\epsilon_0$ is the permittivity in vacuum and the effective polarization absorption coefficient $\chi _{{\text{eff}}}^{\left( 5 \right)} \left( \omega , z \right) $ is given by
\begin{align}\label{chi_eff}
&\chi _{{\text{eff}}}^{\left( 5 \right)} \left( \omega , z \right) \nonumber\\
&= {\left( {\frac{\jmath }{2}} \right)^5}\frac{N \left( z \right) }{{{\epsilon_0}{\hbar ^5}}}\frac{{{\mu _{12}}{\mu _{23}}{\mu _{34}}{\mu _{45}}{\mu _{56}}{\mu _{61}}}}{{{D_2}{D_3}{D_4}\left[ {{D_5}\left( \omega  \right){D_6}\left( \omega  \right) + \frac{{{{\left| {{\Omega _{\text{A}}}} \right|}^2}}}{4}} \right]}}.
\end{align}
For a cold atomic cloud with a spatially varying density, $\chi _{{\text{eff}}}^{\left( 5 \right)} \left( \omega , z \right)$ inherits the density profile $N \left(z \right) $, and also includes the longitudinal variation of the optical field amplitudes due to finite beam waists.

Given ${P^{\left( 5 \right)}}\left( \omega,z  \right)$, the next step is to determine how this polarization generates an output light field (L) that we can actually detect. 
Here we adopt a collinear, paraxial, and slowly-varying-envelope treatment, so that the propagation is reduced to an effective one-dimensional problem along the $z$-axis. This reduction is introduced as a compact model for the RF-to-optical transduction chain. 
Given the present cold-atom configuration with strongly suppressed Doppler and collisional broadening, transit-time effects are not treated here as a separate dominant broadening mechanism~\cite{SWM-2,SWM-2.5,BBR-1,BBR-3}. The residual contribution of such motion-induced effects is absorbed into the effective coherence-linewidth parameters. Therefore, the field evolution satisfies
\begin{align}\label{E_L}
	{E_{\text{L}}}\left( \omega  \right) =& \jmath \frac{{{\omega _{\text{L}}}}}{{2{n_{\text{L}}}c}}{E_{{\text{RF}}}}\left( \omega  \right)
	\nonumber\\
	&\times \int {\chi _{{\text{eff}}}^{\left( 5 \right)}\left( {\omega ,z} \right){E_{\text{P}}}\left( z \right){E_{\text{C}}}\left( z \right){E_{{\text{LO}}}}\left( z \right)E_{\text{A}}^*\left( z \right)} \ dz,
\end{align}
where $n_\text{L}$ is the refractive index of the light field (L) at frequency $\omega_{\text{L}}$ and $c$ is the speed of light.
For compactness, we further introduce an overlap-weighted effective length $L_\text{eff}$ to collect the longitudinal density profile together with the spatial overlap of the participating fields along the propagation axis. Let $N_0 = N\left( 0 \right) $ denote the peak density at the cloud center, and then we define
\begin{equation}\label{L_eff}
{L_{{\text{eff}}}} \triangleq \frac{{\int {\frac{{N\left( z \right)}}{{{N_0}}}{E_{\text{P}}}\left( z \right){E_{\text{C}}}\left( z \right){E_{{\text{LO}}}}\left( z \right)E_{\text{A}}^*\left( z \right)dz} }}{{{E_{\text{P}}}\left( 0 \right){E_{\text{C}}}\left( 0 \right){E_{{\text{LO}}}}\left( 0 \right)E_{\text{A}}^*\left( 0 \right)}},
\end{equation}
where $z=0$ denotes the cloud center; hereafter, we drop this notation for ease of exposition. In this sense, $L_\text{eff}$ is not an arbitrary fitting length, but a normalized geometrical overlap factor that summarizes the axial inhomogeneity of the atomic density and the longitudinal beam-weighting in a closed-form manner.
Then, using (\ref{chi_eff}) and the fact that $\chi _{{\text{eff}}}^{\left( 5 \right)}\left( {\omega ,z} \right) \propto N\left( z \right)$, the integral in (\ref{E_L}) can be written as a compact form
\begin{equation}
{E_{\text{L}}}\left( \omega  \right) = {G_{{\text{opt}}}}\left( \omega  \right){E_{{\text{RF}}}}\left( \omega  \right),
\end{equation}
with the optical RF-to-optical transfer function given by
\begin{equation}\label{G_opt}
{G_{{\text{opt}}}}\left( \omega  \right) = \jmath \frac{{{\omega _{\text{L}}}{L_{{\text{eff}}}}}}{{2{n_{\text{L}}}c}}\chi _{{\text{eff}}}^{\left( 5 \right)}\left( \omega  \right){E_{\text{P}}}{E_{\text{C}}}{E_{{\text{LO}}}}E_{\text{A}}^*.
\end{equation}
By employing the effective polarization absorption coefficient $\chi _{{\text{eff}}}^{\left( 5 \right)}$ shown in (\ref{chi_eff}), one sees explicitly that ${G_{{\text{opt}}}}\left( \omega  \right)$ inherits the same two-pole low-pass structure determined by $D_5 \left( \omega\right) $ and $D_6 \left( \omega \right) $. In the time domain, the output light field can be expressed as a convolution, i.e., ${E_{\text{L}}}\left( t \right) = {g_{{\text{opt}}}}\left( t \right) * {E_\text{RF}}\left( t \right)$, where the impulse response corresponding to the two complex poles at ${ - \left( {{\gamma _{51}} + \jmath {\Delta _{51}}} \right)}$ and ${ - \left( {{\gamma _{61}} + \jmath {\Delta _{61}}} \right)}$ takes the form of
\begin{align}
{g_{{\text{opt}}}}\left( t \right) =& \jmath \frac{{{\omega _{\text{L}}} L_\text{eff}}}{{2{n_{\text{L}}}c}}{\chi _0}{E_{\text{P}}}{E_{\text{C}}}{E_{{\text{LO}}}}E_{\text{A}}^* \nonumber\\
& \times \frac{{{e^{ - \left( {{\gamma _{51}} + \jmath {\Delta _{51}}} \right)t}} - {e^{ - \left( {{\gamma _{61}} + \jmath {\Delta _{61}}} \right)t}}}}{{\left( {{\gamma _{61}} - {\gamma _{51}}} \right) + \jmath \left( {{\Delta _{61}} - {\Delta _{51}}} \right)}}  u\left( t \right),
\end{align}
where $u\left( t \right)$ denotes the unit-step function.

Actually, the output light field  ${E_{\text{L}}}\left( t \right)$ is not measured directly. Instead, we interfere it with the probe beam on the same PD, and use this optical beat to convert the RF information into an electrical current. We first combine the probe field (P) and the output light field (L) on a single PD. The total optical field incident on the diode can be written as
\begin{equation}
E\left( t \right) = {E_{\text{P}}}{e^{\jmath {\omega _P}t}} + {E_{\text{L}}}{e^{\jmath {\omega _L}t}},\left| {{E_\text{L}}} \right| \ll \left| {{E_\text{P}}} \right|,
\end{equation}
which leads to the instantaneous optical power given by
\begin{align}\label{P_opt}
{P_{{\text{opt}}}}\left( t \right) &= \frac{1}{2}n_\text{L}{\epsilon_0}c\int_{{A_{{\text{eff}}}}} {{{\left| {E\left( t \right)} \right|}^2}} d{A_{{\text{eff}}}} \nonumber\\
& \equiv \frac{1}{2}n_\text{L}{\epsilon_0}c{A_{{\text{eff}}}}\left( {{{\left| {{E_{\text{P}}}} \right|}^2} + {{\left| {{E_{\text{L}}}} \right|}^2} + 2\Re \left[ {{E_{\text{P}}}E_{\text{L}}^*{e^{\jmath \Delta \omega t}}} \right]} \right),
\end{align}
where $A_\text{eff}$ is the effective area~\cite{chen-arxiv} and $\Delta \omega  = {\omega _{\text{P}}} - {\omega _{\text{L}}} $ denotes the optical beat frequency close to the RF intermediate frequency. 
Let $R_{\text{pd}}$ denote the PD responsivity (with units of A/W). The resultant photocurrent is
\begin{align}\label{photocurrent}
i\left( t \right) = {R_{{\text{pd}}}}{P_{{\text{opt}}}}\left( t \right) 
 =& \underbrace {\frac{1}{2}n_\text{L}{\epsilon_0}c{R_{{\text{pd}}}}{A_{{\text{eff}}}}\left( {{{\left| {{E_{\text{P}}}} \right|}^2} + {{\left| {{E_{\text{L}}}} \right|}^2}} \right)}_{{I_{{\text{DC}}}}}  \nonumber\\
&+ \underbrace {{R_{{\text{pd}}}}n_\text{L}{\epsilon_0}c{A_{{\text{eff}}}}\Re \left[ {{E_{\text{P}}}E_{\text{L}}^*{e^{\jmath \Delta \omega t}}} \right]}_{{I_{{\text{AC}}}}\left( t \right)},
\end{align}
where $I_{{\text{DC}}}$ collects the slowly varying terms and $I_{{\text{AC}}}$ captures the heterodyne beat current. For notational simplicity, we define a constant $\kappa  \equiv {n_{\text{L}}}{\epsilon_0}c{A_{{\text{eff}}}}$ having units of ${\left[ {{\text{W}} \cdot {{\text{m}}^2}/{{\text{V}}^2}} \right]}$.
Substituting $E_{\text{L}}$ into the beat-current expression yields the frequency-domain baseband current
\begin{equation}
i\left( \omega  \right) = {R_{{\text{pd}}}}\kappa {E_{\text{P}}}{G_{{\text{opt}}}}\left( \omega  \right){E_{{\text{RF}}}}\left( \omega  \right).
\end{equation}
Finally, the photocurrent passes through a transimpedance amplifier or low-noise amplifier (LNA) having a gain of ${G_{{\text{LNA}}}}$, resulting in the receiver output voltage 
\begin{equation}\label{signal_model}
y\left( \omega  \right) = {G_{{\text{LNA}}}}{R_{{\text{pd}}}}\kappa {E_{\text{P}}}{G_{{\text{opt}}}}\left( \omega  \right){E_{{\text{RF}}}}\left( \omega  \right) + n\left( \omega  \right),
\end{equation}
where $n \left( \omega\right) $ represents the aggregate baseband noise observed  at the receiver output. 
It is comprised of contributions from several physical mechanisms, such as blackbody-radiation (BBR)-induced external noise, photon shot noise (PSN) associated with the probe and generated optical fields, quantum projection noise (QPN), PD noise, and so on. A detailed noise budget for SWM-based Rydberg atomic receivers follows directly from the modeling framework in our earlier work~\cite{chen-arxiv}.

At this point, it is important to distinguish between the internally generated atomic coherence, the optical field produced by the SWM process, and the experimentally accessible detected output. Although $\rho_{61} \left( \omega \right) $ and $E_\text{L} \left( \omega \right) $ deliver a microscopic description of the RF-to-optical conversion chain, the actual observable is the detected heterodyne readout $y\left( \omega  \right) $ presented in~\eqref{signal_model}. Therefore, the receiver's transfer characteristics should be defined at the level of the detected output. More specifically, let $\bm{\rho}$ collect the independent density-matrix components of the six-level system, and let  ${\bm{\rho}}_0$ denote the steady-state operating point established by the fields (P, C, LO, and A). Under a weak RF signal $\delta \Omega_{{\text{RF}}} \left( t \right) $, the state variation $\delta {\bm{\rho}} \left( t \right)  $ obeys the linearized dynamics
\begin{equation}
\delta \dot {\bm{\rho}} \left( t \right) = {\mathbf{A}} \delta {\bm{\rho}} \left( t \right) + {\mathbf{b}} \delta {\Omega _{{\text{RF}}}}\left( t \right),
\end{equation}
where ${\mathbf{A}}$ is the Jacobian of the full Liouvillian evaluated at 
${\bm{\rho}}_0$, and $\mathbf{b}$ is the RF-coupling vector. If the detected output variation is denoted by $\delta y \left( t \right) $, then we have $\delta y\left( t \right) = {{\mathbf{c}}^T}\delta {\bm{\rho}} \left( t \right)$, where $\mathbf{c}$ represents the linear projection. Accordingly, the exact response is defined as
\begin{equation}\label{H_exact}
	{H_{{\text{exact}}}}\left( {\jmath \omega } \right) \triangleq \frac{{\delta y\left( {\jmath \omega } \right)}}{{\delta {\Omega _{{\text{RF}}}}\left( {\jmath \omega } \right)}} = {{\mathbf{c}}^T}{\left( {\jmath \omega {\mathbf{I}} - {\mathbf{A}}} \right)^{ - 1}}{\mathbf{b}}.
\end{equation}
This definition makes clear that the exact response of the SWM-based Rydberg atomic receiver is, in general, higher-order and may contain additional modal structure beyond any reduced closed-form expression. The compact response kernel obtained from $\rho_{61} \left( \omega \right) $ should therefore be interpreted as an analytically tractable approximation to the dominant low-frequency behavior of the detected SWM response around the operating point of interest.

\section{Performance Analysis}\label{Sec_III}

Given the baseband signal model presented in (\ref{signal_model}), the key performance metrics of interest are the available 3-dB bandwidth and the linear dynamic range of the SWM-based Rydberg atomic receiver. Since the optical and electronic stages are designed to be nearly flat over the frequency region of interest, the overall RF bandwidth is essentially determined by the atomic response embedded in ${G_{{\text{opt}}}}$. In the following, we first identify how the six-level SWM process gives rise to an effective two-pole low-pass response that sets the 3-dB bandwidth. Based on this understanding, we then discuss the linear dynamic range and the trade-offs between bandwidth and nonlinearity for the proposed SWM architecture.

\subsection{3-dB Bandwidth Analysis}

The instantaneous 3-dB bandwidth considered in this work is defined from the exact response ${H_{{\text{exact}}}}\left( {\jmath \omega } \right)$ presented in~\eqref{H_exact}.{\footnote{Throughout this paper, the optical spectral linewidth or FWHM refers to a static spectroscopic feature obtained from detuning scans, whereas $f_{3\text{dB}}$ denotes the RF-to-optical transduction bandwidth extracted from the detected baseband response versus modulation frequency. The measurement bandwidth, or noise-equivalent bandwidth, may be set equal to the 3-dB bandwidth for a full-passband sensitivity estimate; however, it is conceptually distinct from the static optical FWHM.}} Specifically, we defined the 3-dB bandwidth $f_\text{3dB}$ through the operational condition 
\begin{equation}\label{3dB_condition}
\left| {{H_{{\text{exact}}}}\left( {\jmath 2\pi {f_{{\text{3dB}}}}} \right)} \right| = \frac{{\left| {{H_{{\text{exact}}}}\left( 0 \right)} \right|}}{{\sqrt 2 }}.
\end{equation}
This definition is also consistent with the communication-oriented notion that the usable bandwidth is extracted from the detected beat-response versus modulation frequency~\cite{ref_1_yang2024}.

To gain an analytical insight, we next introduce a reduced-order approximation around the operating point. Recall that the detected SWM response can be decomposed into a lower-level factor associated with ${D_2}\left( \omega  \right)$, ${D_3}\left( \omega  \right)$, and ${D_4}\left( \omega  \right)$, as well as a higher-level dressed kernel associated with ${D_5}\left( \omega  \right)$, ${D_6}\left( \omega  \right)$, and $ \left|  \Omega_{{\text{A}}} \right| $. By combining $\rho_{61}$ in~\eqref{rho_61} with the formulation in~\eqref{D2_D3_D4}, the dominant low-frequency envelope of the detected SWM response, in the near-resonant weak-signal regime of interest, can be restructured in an approximate form
\begin{equation}\label{H_approx}
{H_{{\text{approx}}}}\left( {\jmath \omega } \right) \approx \tilde G\left( 0 \right)\mathfrak{P}\left( \omega \right)\frac{1}{{{D_5}\left( \omega  \right){D_6}\left( \omega  \right) + {{\left| {{\Omega _{\text{A}}}} \right|}^2}/4}},
\end{equation}
where ${{\tilde G}_0} = 1 / {\left( {D_2^{\left( 0 \right)}D_3^{\left( 0 \right)}D_4^{\left( 0 \right)}} \right)}$ denotes the constant part of the lower-level factor, and $ \mathfrak{P}\left( \omega  \right) $ is the corresponding normalized frequency-dependent correction term, which satisfies  $ \mathfrak{P}\left( \omega  \right) \approx 1 $. Appendix~\ref{appendix_error_analysis} provides a detailed error analysis of approximating the lower-stage factor as frequency-flat, namely $\mathfrak{P}\left( \omega  \right) \approx 1$. The remaining approximately constant contributions from the optical readout chain and the electronic gain are absorbed into $\tilde G\left( 0 \right)$.

Under this reduced-order approximation, the higher-level dressed kernel determines the characteristic low-frequency roll-off. For operation near the intended working point, where ${\Delta _{51}} \approx {\Delta _{61}} \approx 0$ and the detected SWM response remains monotonic around DC, the denominator in~\eqref{H_approx} in the standard second-order form
\begin{equation}
{D_5}\left( \omega  \right){D_6}\left( \omega  \right) + \frac{{{{\left| {{\Omega _{\text{A}}}} \right|}^2}}}{4} = \left( {\jmath\omega - {\lambda _ + }} \right)\left( {\jmath\omega - {\lambda _ - }} \right),
\end{equation}
with two poles $\lambda_\pm$ given by
\begin{equation}
{\lambda _ \pm } =  - \frac{{{\gamma _{51}} + {\gamma _{61}}}}{2} \pm \frac{1}{2}\sqrt {{{\left( {{\gamma _{51}} - {\gamma _{61}}} \right)}^2} - {{\left| {{\Omega _{\text{A}}}} \right|}^2}} .
\end{equation}
The real parts of these poles control how fast the RF-induced excitation decays in time. We therefore define the corresponding decay rates
\begin{equation}\label{gamma_pm}
{\gamma _ \pm } \equiv  - \Re \left\{ {{\lambda _ \pm }} \right\} ,
\end{equation}
which have units of rad/s. Intuitively, $\gamma_-$ characterizes the slow model (longer-lived response) and $\gamma_+$ captures the fast mode. Note that the physical meaning of  ${\gamma _ \pm }$ is not the optical linewidth in the sense of FWHM. Instead, they represent the inverse time constants, or equivalently, the half-power angular frequencies of two effective first-order modes that together constitute the overall second-order response.
In the overdamped regime $ {{\Omega _A}}  < \left| {{\gamma _{51}} - {\gamma _{61}}} \right|$, both $\lambda_+$ and $\lambda_-$ are real and negative, so the SWM chain behaves as a monotonic two-pole low-pass system.

Then, normalizing the optical transfer function to its DC gain, the magnitude response can be given by
\begin{equation}
\frac{{{{\left| {{H_{{\text{approx}}}}\left( {\jmath \omega } \right)} \right|}^2}}}{{{{\left| {{H_{{\text{approx}}}}\left( 0 \right)} \right|}^2}}} = \frac{1}{{\left( {1 + {\omega ^2}/\gamma _ + ^2} \right)\left( {1 + {\omega ^2}/\gamma _ - ^2} \right)}}.
\end{equation}
The 3-dB bandwidth $\omega_\text{3dB}$ of the dominant low-frequency envelope is obtained from
\begin{equation}\label{H_approx_3dB}
\frac{{{{\left| {{H_{{\text{approx}}}}\left( {\jmath {\omega _{3{\text{dB}}}}} \right)} \right|}^2}}}{{{{\left| {{H_{{\text{approx}}}}\left( 0 \right)} \right|}^2}}} = \frac{1}{2},
\end{equation}
which leads to 
\begin{equation}
\left[ {1 + {{\left( {\frac{{{\omega _{3{\text{dB}}}}}}{{{\gamma _ + }}}} \right)}^2}} \right]\left[ {1 + {{\left( {\frac{{{\omega _{3{\text{dB}}}}}}{{{\gamma _ - }}}} \right)}^2}} \right] = 2.
\end{equation}
Solving this quadratic equation yields a closed-form expression
\begin{equation}\label{omega_3dB}
\omega _{{\text{3dB}}}^2 = \frac{{ - \left( {\gamma _ + ^2 + \gamma _ - ^2} \right) + \sqrt {{{\left( {\gamma _ + ^2 + \gamma _ - ^2} \right)}^2} + 4\gamma _ + ^2\gamma _ - ^2} }}{2}.
\end{equation}
Therefore, the 3-dB bandwidth $\omega_{{\text{3dB}}}$ in~\eqref{omega_3dB} should be interpreted as a closed-form approximation to the cutoff of the dominant low-frequency envelope, rather than as a globally exact bandwidth law for the full six-level SWM-based receiver. In this sense, $\gamma_{51}$, $\gamma_{61}$, and $\Omega_{{\text{A}}}$ should not be understood as uniquely or universally determining the exact detected bandwidth in all regimes; instead, they primarily determine the characteristic roll-off within the reduced-order model and the near-resonant weak-signal regime examined here. 

Our reduced-order model differs from the work in~\cite{ref_1_yang2024}, where the instantaneous bandwidth is defined directly from the detected beat response between the generated sidebands and the probe carrier, rather than from a reduced envelope-level approximation. In that setting, the bandwidth enhancement arises from an imbalance between the two four-wave mixing pathways: the negative-sideband contribution is selectively enhanced under strong coupling, which produces a finite-frequency gain peak and reshapes the measured modulation response. As a stark contrast, our reduced-order model characterizes the dominant low-frequency roll-off of the RF-to-optical transduction response around the chosen operating point. The two results therefore refer to different observables and different levels of approximation, and are complementary rather than contradictory.

\subsection{Sensitivity Analysis}\label{sec_sensitivity_analysis}
We now quantify the electric-field sensitivity of the SWM-based Rydberg atomic receiver. To place different noise sources on a common footing, we adopt the NEF~\cite{QS-32,QS-34}. NEF is defined as the minimum input RF field amplitude that yields an output $ \text{SNR} = 1$ over an effective noise bandwidth $B${\footnote{Here, $B$ denotes the measurement (or noise-equivalent) bandwidth used to integrate noise power, i.e., the effective bandwidth of the baseband filtering employed in the sensitivity readout. It is therefore not necessarily identical to the receiver's small-signal 3-dB bandwidth characterized by $\omega_{\text{3dB}}$. For a fair, order-of-magnitude sensitivity evaluation and for comparing different receiver configurations under the same full-passband operation, we may choose $B \approx f_\text{3dB}$.}}:
\begin{equation}
	{\text{NEF}}  \triangleq  \frac{{{{\left| {{E_{{\text{RF}}}}} \right|}_{\min }}}}{{\sqrt B }} \ \left( {{\text{V/m/}}\sqrt {{\text{Hz}}} } \right) ,
\end{equation}
where $ \left| E_{\text{RF}} \right| _{\min} $ denotes the minimum detectable field.

We then investigate the noise sources and their NEFs, considering one extrinsic contribution (free-space BBR noise) and several intrinsic contributions, in which each term is expressed as a NEF referred to the RF input.
\begin{itemize}
\item Extrinsic Noise Induced by Black-Body Radiation (BBR): As established in our prior work~\cite{chen-arxiv}, the external field noise floor ${\text{NE}}{{\text{F}}_{{\text{ex}}}}$ is modeled as
\begin{equation}
{\text{NE}}{{\text{F}}_{{\text{ex}}}} = \sqrt {\frac{{16\pi f_{{\text{RF}}}^2}}{{3{\epsilon _0}{c^3}}} \Theta \left( {{f_{{\text{RF}}}},T} \right) } ,
\end{equation}
where $\Theta \left( {{f_{{\text{RF}}}},T} \right)$ represents a modified version of the Callen-Welton law~\cite{Welton_1951}, $T$ denotes the ambient physical temperature, $f_{{\text{RF}}}$ is the RF signal frequency, $\epsilon_{0}$ is the vacuum permittivity, and $c$ is the speed of light.

\item \textbf{Quantum Projection Noise (QPN):} For an ensemble of $N_\text{atoms}$ atoms, the SQL-limited field noise is given by
\begin{equation}
{\text{NE}}{{\text{F}}_{{\text{QPN}}}} = \frac{\hbar }{{{\wp _{{\text{RF}}}}\sqrt {{N_{{\text{atoms}}}}} {T_2}}},
\end{equation}
where ${{\wp _{{\text{RF}}}}}$ represents the RF dipole matrix element and $T_2$ is the coherence time.

\item \textbf{Photon Shot Noise (PSN):} From~(\ref{photocurrent}), DC photocurrent $I_\text{DC}$ can be readily obtained. The corresponding power spectral density of PSN is given by $2 e I_\text{DC}$, where $e$ denotes the elementary charge. Referring this noise back to the SWM system's RF input yields
\begin{equation}
{\text{NE}}{{\text{F}}_{{\text{PSN}}}} = \frac{{\sqrt {2e{I_{{\text{DC}}}}} }}{{{R_{{\text{pd}}}}\kappa {E_{\text{P}}}\left| {{G_{{\text{opt}}}}\left( 0 \right)} \right|}}.
\end{equation}

\item \textbf{Laser Relative Intensity Noise (RIN):} RIN characterizes classical amplitude fluctuations of the laser power. In the optical power domain, the power spectral density, denoted by ${S_{P,{\text{RIN}}}}$, is given by 
\begin{equation}
{S_{P,{\text{RIN}}}} = {\left| {{\bar{P}_{{\text{opt}}}}} \right|^2}{S_{{\text{RIN}}}},
\end{equation}
where ${{\bar{P}_{{\text{opt}}}}}$ takes the average optical power incident on the photodetector (i.e., $P_\text{out} \left( t \right) $ in~(\ref{P_opt})). When expressed in the photocurrent domain, the power spectral density, denoted by ${S_{i{\text{,RIN}}}}$, is in the form of
\begin{equation}
{S_{i,{\text{RIN}}}} = R_{{\text{pd}}}^2{\left| {{\bar{P}_{{\text{opt}}}}} \right|^2}{S_{{\text{RIN}}}}.
\end{equation}
We then obtain the RF-input-referred NEF
\begin{equation}
{\text{NE}}{{\text{F}}_{{\text{RIN}}}} = \frac{{\left| {{{\bar P}_{{\text{opt}}}}} \right|\sqrt {{S_{{\text{RIN}}}}} }}{{\kappa {E_{\text{P}}}\left| {{G_{{\text{opt}}}}\left( 0 \right)} \right|}}.
\end{equation}
In practice, $S_{{\text{RIN}}}$ can be taken from datasheets~\cite{RIN_1,RIN_2} or measured.

\item \textbf{Thermal Noise (TN):} The spectral power of TN over bandwidth $B$ is modeled as
\begin{equation}
S_{\text{TN}} = \left\{ \begin{gathered}
	4{k_{\text{B}}}TB, \ {\text{TIA front-end}}, \hfill \\
	F{k_{\text{B}}}TB, \, {\text{LNA front-end}}, \hfill \\ 
\end{gathered}  \right.
\end{equation}
where $F$ is the noise factor. By referring this voltage noise to the RF input, we arrive at the noise field given by
\begin{equation}
{\text{NE}}{{\text{F}}_{{\text{TN}}}} = \frac{{\sqrt {{S_{{\text{TN}}}}} }}{{{G_{{\text{LNA}}}}{R_{{\text{pd}}}}\kappa {E_{\text{P}}}\left| {{G_{{\text{opt}}}}\left( 0 \right)} \right|}}.
\end{equation}
\end{itemize}
Since QPN associated with the atoms and optical fields are typically correlated, and the internal noise floor is given by
\begin{equation}
	\resizebox{\linewidth}{!}{$
{\text{NE}}{{\text{F}}_{{\text{in}}}} = \sqrt {{\text{NEF}}_{{\text{QPN}}}^2 + 2r{\text{NE}}{{\text{F}}_{{\text{QPN}}}}{\text{NE}}{{\text{F}}_{{\text{PSN}}}} + {\text{NEF}}_{{\text{PSN}}}^2 + {\text{NEF}}_{{\text{RIN}}}^2 + {\text{NEF}}_{{\text{TN}}}^2} ,
$}
\end{equation}
where $r$ represents the statistical correlation between ${{\text{NE}}{{\text{F}}_{{\text{QPN}}}}}$ and ${{\text{NE}}{{\text{F}}_{{\text{PSN}}}}}$, and the correlation coefficient $r$ can be computed based on the fact in~\cite[Eq.~(S16)]{QS-34}. 
Ultimately, we obtain the achievable sensitivity for the SWM-based Rydberg atomic receiver of 
\begin{equation}
\text{NEF}_\text{tot} = 
\sqrt {{\text{NEF}}_{{\text{ex}}}^2 + {\text{NEF}}_{{\text{in}}}^2}.
\end{equation}

\subsection{Linear Dynamic Range}
In addition to bandwidth, the receiver has to remain sufficiently linear over an appropriate  range of input amplitudes; otherwise, strong signals will induce gain compression or generate in-band distortion that degrades detection and demodulation. For the SWM-based Rydberg atomic receiver, we characterize the linear dynamic range using two widely adopted metrics: P1dB and IIP3, which are briefly summarized as follows.
\begin{itemize}
\item \textbf{1-dB compression point (P1dB):} P1dB quantifies the onset of saturation under a single-tone excitation. As the input amplitude increases, the output no longer scales proportionally with the input, while the small-signal gain starts to drop. The P1dB point is defined as the input level at which the gain is reduced by 1 dB compared with the extrapolated linear response. In simple terms, a larger P1dB indicates that the receiver preserves linear amplification over a wider input range.

\item  \textbf{Input-referred third-order intercept point  (IIP3):}  IIP3 quantifies third-order nonlinear distortion under a two-tone excitation. When a pair of closely-spaced tones at $\omega_1$ and $\omega_2$ are applied, third-order nonlinearity generates inter-modulation products at $2\omega_1 - \omega_2$ and $2\omega_2 - \omega_1$, which typically fall back into (or near) the signal band and thus are particularly harmful for wideband or multi-carrier signals. A higher IIP3 implies better linearity and weaker in-band third-order inter-modulation distortion (IMD3) for the same input level.

\end{itemize}
In our atomic system, the ``input amplitude" is the RF Rabi frequency $\Omega_{{\text{RF}}}$, which is proportional to the incident RF electric-field amplitude. The ``output" can be chosen as $\rho_{61} \left( \omega \right) $ (or equivalently, the detected baseband voltage $y$), because the optical readout and electronic chain are assumed linear and therefore introduce only a constant scaling. With these definitions, we next derive analytical expressions for P1dB and IIP3.

\subsubsection{P1dB}
We first consider a single-tone RF drive at angular frequency  $\omega$. The gain from the RF Rabi frequency to $\rho_{61} \left( \omega \right) $ is defined as
\begin{equation}\label{tmp_gain}
G\left( {\omega ,{\Omega _{{\text{RF}}}}} \right) \equiv \frac{{{\rho _{61}}\left( \omega  \right)}}{{{\Omega _{{\text{RF}}}}}}.
\end{equation}
In the weak-nonlinearity regime, the dependence of the detected SWM response on the input RF amplitude can be locally approximated by an odd-order expansion, which is given by
\begin{equation}\label{third_order}
	{\rho _{61}}\left( \omega  \right) \approx {H_1}\left( \omega  \right){\Omega _{{\text{RF}}}} + \frac{3}{4}{H_3}\left( \omega  \right)\Omega _{{\text{RF}}}^3.
\end{equation}
Here, ${H_1}\left( \omega  \right) \triangleq {\left. {\frac{{\partial {\rho _{61}}\left( \omega  \right)}}{{\partial {\Omega _{{\text{RF}}}}}}} \right|_{{\Omega _{{\text{RF}}}} = 0}} $ and ${H_3}\left( \omega  \right) \triangleq {\left. {\frac{2}{9}\frac{{{\partial ^3}{\rho _{61}}\left( \omega  \right)}}{{\partial \Omega _{{\text{RF}}}^3}}} \right|_{{\Omega _{{\text{RF}}}} = 0}} $ denote the linear and the third-order local response coefficients, respectively, in the expansion of $\rho_{61}$.{\footnote{Note that the fifth-order nature of SWM refers to the order of the generated polarization with respect to the participating EM fields, whereas the expansion here is a local amplitude expansion with respect to the input RF drive $\Omega_{{\text{RF}}}$ around the selected operating point.}} Equivalently, they can be interpreted as the first-order and third-order coefficients of the local Taylor expansion with respect to $\Omega_{{\text{RF}}}$. Hence, the expansion in~\eqref{third_order} should be understood as a local third-order approximation introduced for the analytical interpretation of P1dB and IIP3, rather than as an intuition that higher-order nonlinear terms are absent in the full six-level SWM dynamics.
Higher-order odd terms, such as fifth-order corrections, are neglected here only under the weak-signal, weak-nonlinearity assumption~\cite{SWM-2,SWM-2.5}.
Under this local third-order model, the first nonzero odd nonlinear correction is sufficient to capture the onset of gain compression and the  third-order inter-modulation products.
Then, the gain in~(\ref{tmp_gain}) can be recast as
\begin{equation}
G\left( {\omega ,{\Omega _{{\text{RF}}}}} \right) \approx {H_1}\left( \omega  \right)\left[ {1 + \frac{3}{4}\frac{{{H_3}\left( \omega  \right)}}{{{H_1}\left( \omega  \right)}}\Omega _{{\text{RF}}}^2} \right].
\end{equation}
For most operating points of interest, the real part of ${{H_3}\left( \omega  \right)/{H_1}\left( \omega  \right)}$ is negative, corresponding to gain saturation. We define
\begin{equation}
\alpha \left( \omega  \right) \equiv \Re \left\{ {\frac{{{H_3}\left( \omega  \right)}}{{{H_1}\left( \omega  \right)}}} \right\} < 0.
\end{equation}
Normalizing to the gain $G \left( \omega, 0 \right) = H_1 \left( \omega\right)  $
and linearizing the magnitude for $\Omega_{{\text{RF}}}$, we obtain
\begin{equation}
\frac{{G\left( {\omega ,{\Omega_{{\text{RF}}}}} \right)}}{{G\left( {\omega ,0} \right)}}  \approx 1 + \frac{3}{4}\alpha \left( \omega  \right)\Omega _{{\text{RF}}}^2.
\end{equation}
The P1dB is defined as the input level ${\Omega _{{\text{P1dB}}}}$ at which the gain drops by 1~dB relative to the small-signal value:
\begin{align}
& 20{\log _{10}}\frac{{\left| {G\left( {\omega ,{\Omega _{{\text{P1dB}}}}} \right)} \right|}}{{\left| {G\left( {\omega ,0} \right)} \right|}} =  - 1{\text{ dB}} \nonumber\\
\Rightarrow & \frac{{G\left( {\omega ,{\Omega _{{\text{P1dB}}}}} \right)}}{{G\left( {\omega ,0} \right)}} \approx 1 + \frac{3}{4}\alpha \left( \omega  \right)\Omega _{{\text{P1dB}}}^2 = {10^{ - \frac{1}{{20}}}}.
\end{align}
Solving for ${{\Omega _{{\text{P1dB}}}}}$ yields
\begin{equation}
{\Omega _{{\text{P1dB}}}}\left( \omega  \right) \approx \sqrt {\frac{4}{3}\frac{{{{10}^{ - 1/20}} - 1}}{{\alpha \left( \omega  \right)}}} ,\alpha \left( \omega  \right) < 0 .
\end{equation}
This expression explicitly reveals how the third-order coefficient ${H_3}\left( \omega  \right)$, through $\alpha \left( \omega  \right)$, controls the onset of gain compression: a smaller $\alpha \left( \omega  \right)$ corresponds to a larger the P1dB and, consequently, an expanded linear input range.

\subsubsection{IIP3}

To characterize the receiver linearity under multi-tone or modulated RF signals, we next consider a standard two-tone test. The RF Rabi frequency is taken as an equal-amplitude two-tone signal ${\Omega _{{\text{RF}}}}\left( t \right) = \Omega \cos {\omega _1}t + \Omega \cos {\omega _2}t$, with $\omega_1$ and $\omega_2$ chosen close to each other and within the signal band. Substituting this drive into the third-order expansion in (\ref{third_order}) and keeping terms up to third order, we obtain the components of $\rho_{61} \left( \omega \right) $ at the 
fundamental and third-order inter-modulation frequencies. Specifically, at the fundamental frequency $\omega_1$, the response is given by
\begin{equation}
{\rho _{61}}\left( {{\omega _1}} \right) \approx {H_1}\left( \omega  \right){\Omega _{{\text{RF}}}} + \frac{3}{4}{H_3}\left( \omega  \right)\Omega _{{\text{RF}}}^3,
\end{equation}
while at the third-order inter-modulation frequency $2\omega_1 - \omega_2$, we have
\begin{equation}
{\rho _{61}}\left( {2{\omega _1} - {\omega _2}} \right) \approx \frac{3}{4}{H_3}\left( \omega  \right)\Omega _{{\text{RF}}}^3.
\end{equation}
The IMD3 level (in dBc) is defined as the ratio between the inter-modulation tone and the fundamental tone at the same output node:
\begin{align}\label{IMD3}
{\text{IMD3}}\left( {{\text{dBc}}} \right) = 20{\log _{10}}\left| {\frac{3}{4}\frac{{{H_3}\left( \omega  \right)}}{{{H_1}\left( \omega  \right)}}\Omega _{{\text{RF}}}^2} \right|.
\end{align}
The IIP3 is obtained by extrapolating these two trends and finding their intersection. At the intercept, the magnitude of the fundamental and IMD3 tone are equal, which delivers
\begin{align}
&\left| {{H_1}\left( \omega  \right)} \right|{\Omega _{{\text{IIP3}}}} \approx \frac{3}{4}\left| {{H_3}\left( \omega  \right)} \right|\Omega _{{\text{IIP3}}}^3 \nonumber\\
\Rightarrow &{\Omega _{{\text{IIP3}}}}\left( \omega  \right) \approx \sqrt {\frac{4}{3}\left| {\frac{{{H_1}\left( \omega  \right)}}{{{H_3}\left( \omega  \right)}}} \right|} .
\end{align}
A higher IIP3 means that the third-order distortion remains much smaller than the desired signal over a wider input range, which is particularly important for wideband or multi-carrier RF operation. Together with the P1dB analyses above, the expression for both ${\Omega _{{\text{P1dB}}}}$ and ${\Omega _{{\text{IIP3}}}}$ provide a compact characterization of the linear dynamic range of the SWM-based Rydberg atomic receiver, and make clear how it is determined by the frequency-dependent coefficients ${H_1}\left( \omega  \right)$ and ${H_3}\left( \omega  \right)$ of the underlying atomic response.

\section{Simulation Results}\label{Sec_IV_simulation}
\subsection{Parameter Configuration}
We numerically evaluate the six-level SWM-based Rydberg atomic receiver by employing the QuTiP toolkit~\cite{QuTip}, where the parameters are taken from~\cite{SWM-2,SWM-2.5} to ensure physical plausibility. Specifically, energy levels of a $^{87}\text{Rb}$ atom coupled by six nearly-resonant optical fields are given by: $ \left| 1 \right\rangle  = \left| {5{S_{1/2}},F = 2,{m_F} = 2} \right\rangle  $, $\left| 2 \right\rangle  = \left| {5{P_{3/2}},F = 3,{m_F} = 3} \right\rangle $, $\left| 3 \right\rangle  = \left| {30{D_{3/2}},{m_J} = 1/2} \right\rangle $, $\left| 4 \right\rangle  = \left| {31{P_{3/2}},{m_J} = 1/2} \right\rangle $, $\left| 5 \right\rangle  = \left| {30{D_{5/2}},{m_J} = 1/2} \right\rangle $, and $\left| 6 \right\rangle  = \left| {5{P_{3/2}},F = 2,{m_F} = 1} \right\rangle $.
The SWM atomic system considered is driven by five near-resonant fields, i.e., probe~(P), coupling~(C), auxiliary~(A), local oscillator~(LO), and the RF signal, whose Rabi frequencies are configured as ${\Omega _{\text{P}}}/2\pi  = 1.14~{\text{MHz}}$, ${\Omega _{\text{C}}}/2\pi  = 9.0~{\text{MHz}}$, ${\Omega _{\text{A}}}/2\pi  = 6.2~{\text{MHz}}$, ${\Omega _{{\text{LO}}}}/2\pi  = 7.5~{\text{MHz}}$, and ${\Omega _{{\text{RF}}}}/2\pi  = 1.0~{\text{MHz}}$, respectively. Accordingly, the coherence dephasing rate $\gamma_{ij}$ applied can be given by{\color{blue} 
\footnote{Here $\gamma_{21}$ and $\gamma_{61}$ represent optical-linewidth-scale coherence parameters involving the excited states $\left| 2 \right\rangle$  and $\left| 6 \right\rangle $, whereas $\gamma_{31}$, $\gamma_{41}$, and $\gamma_{51}$ correspond to the narrower coherences involving the Rydberg manifold. Given that $\left| 6 \right\rangle $ represents the excited state $5{P_{3/2}},F = 2,m = 1$, rather than a Rydberg state, $\gamma_{61}$ should not be interpreted as a Rydberg-state dephasing rate.}}: ${\gamma _{21}} / 2\pi  = 6.1~{\text{MHz}}$, ${\gamma _{31}}/ 2\pi  = 50~{\text{kHz}}$, ${\gamma _{41}} / 2\pi  = 80~{\text{kHz}}$, ${\gamma _{51}} / 2\pi  = 129~{\text{kHz}}$, and ${\gamma _{61}} / 2\pi  = 6.1~{\text{MHz}}$. The electric dipole matrix elements of the atomic transitions presented in Fig.~\ref{SWM_model}(a) are, in Hartree atomic units, $\left| {{\mu _{12}}} \right| = 2.99$, $\left| {{\mu _{23}}} \right| = 0.00914$, $\left| {{\mu _{34}}} \right| = 211$, $\left| {{\mu _{45}}} \right| = 387$, $\left| {{\mu _{56}}} \right| = 0.0138$, and $\left| {{\mu _{61}}} \right| = 1.22$~\cite{SWM-2.5}.

In more detail, the atomic cloud is Gaussian distributed, with a $1/e^2$ radius of~1.85~mm, a peak atomic density of $N_0 \sim 2\times 10^{10}~\text{cm}^{-3}$, and a temperature of approximately $70~\mu\text{K}$. This volume has a diameter of $\sim 2\times 25~\mu\text{m}$ and a length of $\sim2\times1.85~\text{mm}$. The probe (P), coupling (C), and auxiliary (A) beam lasers are focused onto the atomic cloud with $1/e^2$ radius of 25~$\mu\text{m}$, 54~$\mu\text{m}$, and 45~$\mu\text{m}$, respectively. The beams for both coupling~(C) and auxiliary~(A) are derived from a single 482~nm laser, while that of the probe~(P) field comes from a 780~nm laser. The laser intensities are approximately $20~\mu\text{W}$ for the probe and $17~\text{mW}$ for each of the coupling and auxiliary beams. In the microwave domain, the LO and RF fields are applied at ${f_{{\text{LO}}}} = 84.18~{\text{GHz}}$ and ${f_{{\text{RF}}}} = 83.72~{\text{GHz}}$, respectively. The remaining parameters are specified as follows: $\epsilon_0 = 8.854\times 10^{-12}~\text{F/m}$, $G_\text{LNA} = 20~\text{dB}$, and $R_\text{pd} = 0.55~\text{A/W}$.


\subsection{Performance Evaluation}
\subsubsection{3-dB Bandwidth}

\begin{figure}[t]
	\centering
	\includegraphics[width=0.48\textwidth]{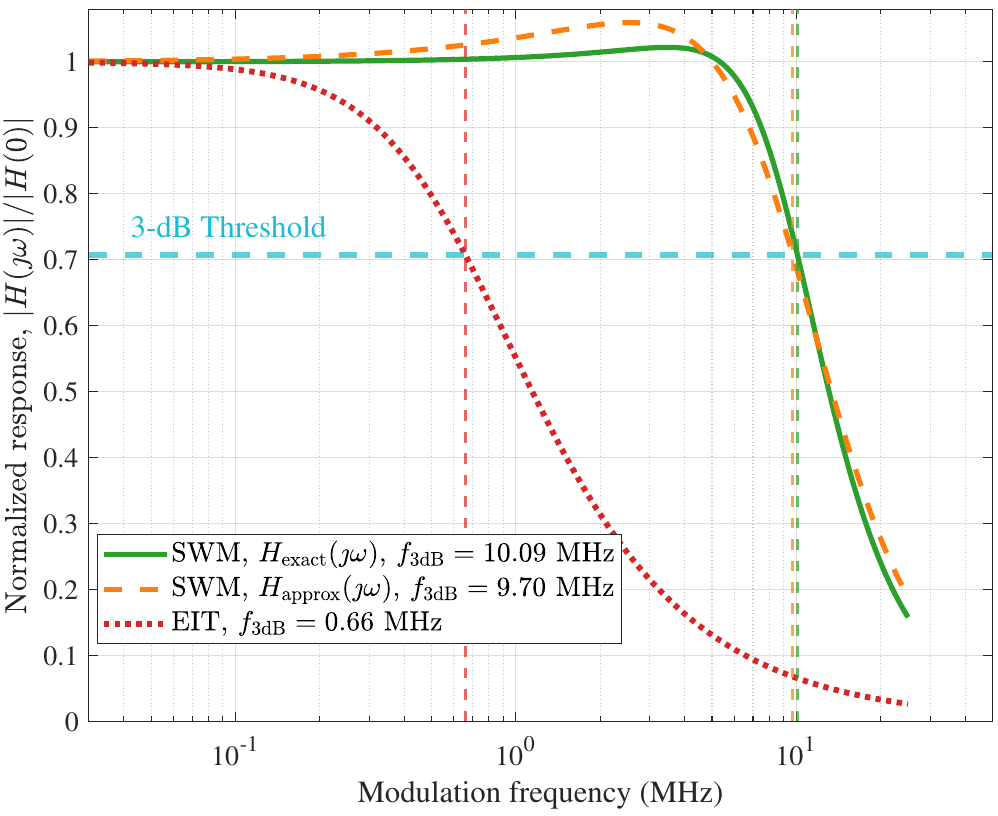}
	\caption{Normalized response and 3-dB bandwidth for both SWM-based and EIT-based Rydberg atomic receivers.} \label{3dB_illustration}
\end{figure}

Fig.~\ref{3dB_illustration} plots the normalized response, ${\left| {H\left( {\jmath \omega } \right)} \right|/\left| {H\left( 0 \right)} \right|}$, for the SWM-based Rydberg atomic receiver with both the exact response $H_\text{exact} \left( \jmath \omega \right) $ formulated in~\eqref{H_exact} and the reduced-order two-pole approximation presented in~\eqref{H_approx}, together with the four-level EIT benchmark. The instantaneous 3-dB bandwidth here is defined based on the detected beat response versus modulation frequency, i.e., the definition in~\eqref{3dB_condition}. In this case, the exact SWM response yields $f_\text{3dB} = 10.09~\text{MHz}$, while the closed-form two-pole approximation gives $f_\text{3dB} = 9.70~\text{MHz}$, indicating close agreement in both cutoff location and dominant low-frequency roll-off. By contrast, the four-level EIT benchmark reaches the same 3-dB threshold at only approximately 0.66~MHz. Thus, the SWM scheme increases the 3-dB bandwidth by about one order of magnitude relative to EIT.

\begin{figure}[!t]
	\centering
	\begin{subfigure}[b]{\linewidth}
		\centering
		\includegraphics[width=0.98\linewidth]{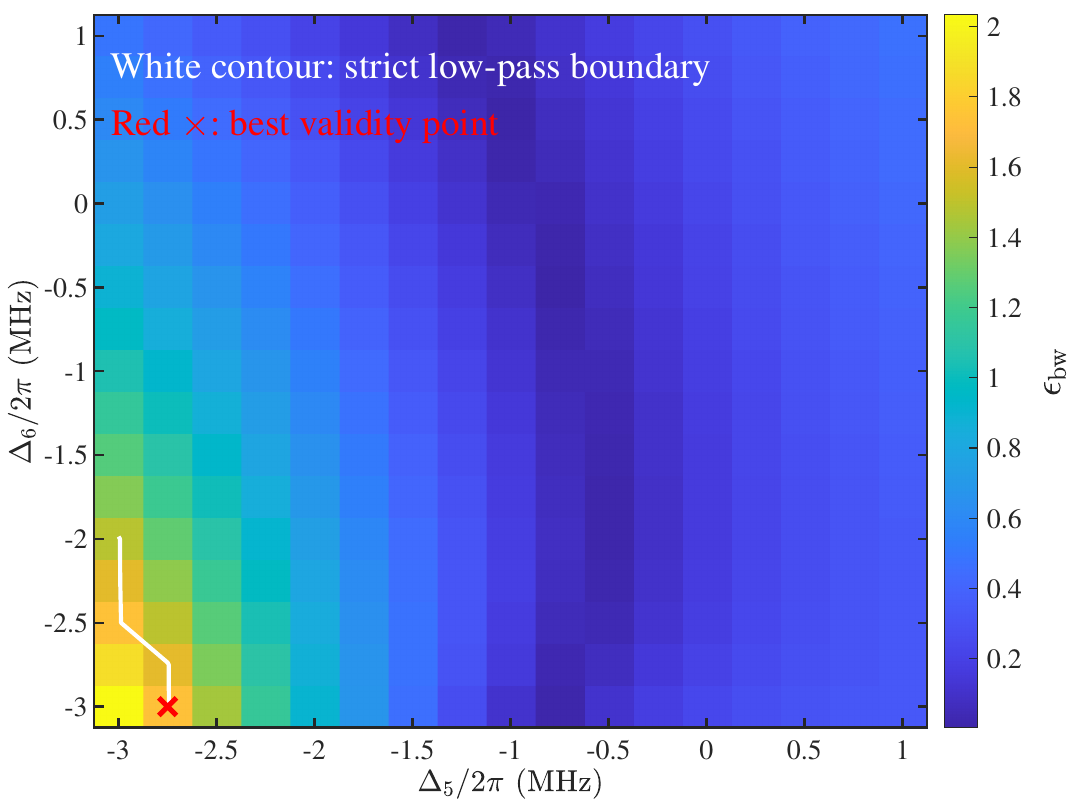}
		\caption{}
		\label{new_fig3_epsilon_bw}
	\end{subfigure}
	\hfill
	\begin{subfigure}[b]{\linewidth}
		\centering
		\includegraphics[width=0.98\linewidth]{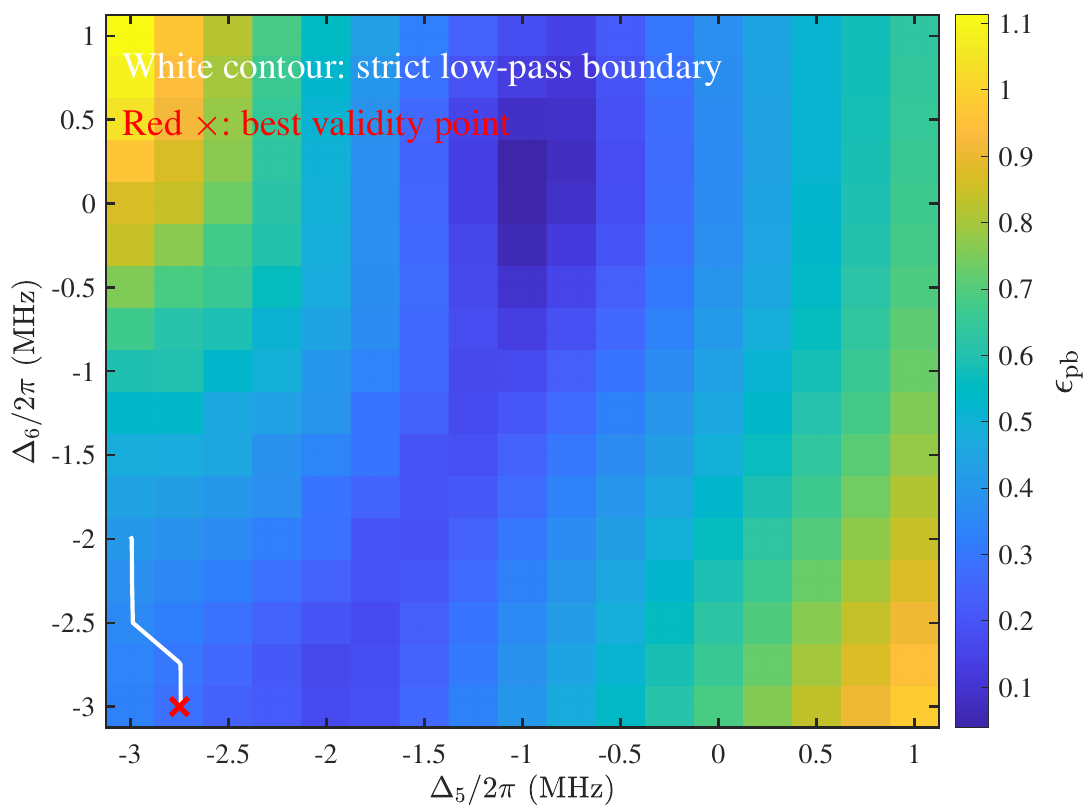}
		\caption{}
		\label{new_fig3_epsilon_pb}
	\end{subfigure}	
	\caption{Approximation validity range over the higher-level operating-point plane $\left( \Delta_5, \Delta_6 \right) $: (a) Bandwidth relative error $\epsilon_\text{bw}$. (b) Passband shape error $\epsilon_{{\text{pb}}}$.}
	\label{new_fig3_epsilon}
\end{figure}

To clarify the validity range of the proposed two-pole approximation in~\eqref{H_approx}, Fig.~\ref{new_fig3_epsilon} maps the approximation error over the higher-level operating-point plane $\left( \Delta_5, \Delta_6 \right) $. Specifically, Fig.~\ref{new_fig3_epsilon_bw} illustrates the bandwidth relative error, ${\epsilon_{{\text{bw}}}}$, which is defined by
\begin{equation}
{\epsilon_{{\text{bw}}}} = \frac{{\left| {f_{{\text{3dB}}}^{{\text{exact}}} - f_{{\text{3dB}}}^{{\text{approx}}}} \right|}}{{f_{{\text{3dB}}}^{{\text{exact}}}}},
\end{equation}
while  Fig.~\ref{new_fig3_epsilon_pb} shows the passband-shape error ${\epsilon_{{\text{pb}}}}$ given by
\begin{equation}
{\epsilon_{{\text{pb}}}} = \left| {\frac{{\left| {{H_{{\text{exact}}}}\left( {\jmath \omega } \right)} \right|}}{{\left| {{H_{{\text{exact}}}}\left( 0 \right)} \right|}} - \frac{{\left| {{H_{{\text{approx}}}}\left( {\jmath \omega } \right)} \right|}}{{\left| {{H_{{\text{approx}}}}\left( 0 \right)} \right|}}} \right|.
\end{equation}
The white contour marks the boundary of the strict low-pass region, while the red cross indicates the best-validity point identified in the scan. The heat map in Fig.~\ref{new_fig3_epsilon} demonstrates that the accuracy of $H_\text{approx} \left( \jmath \omega \right) $ depends much more strongly on the higher-level operating point $\left( \Delta_5, \Delta_6 \right) $ than on a single scalar bandwidth parameter alone. More specifically, near the red-cross region, both the bandwidth relative error $\epsilon_{{\text{bw}}}$ and the passband-shape error $\epsilon_{{\text{pb}}}$ are small, whereas moving away from this region leads to a systematic deterioration of the approximation. This confirms that the practical validity of the proposed two-pole approximation is determined by the selected higher-level operating point and is  also consistent with the Appendix~\ref{appendix_error_analysis} condition that the lower-level factor must remain approximately frequency-flat over the passband.

\subsubsection{3-dB Bandwidth vs. Rabi Frequency}

\begin{figure*}[t]
	\centering
	\includegraphics[width=0.92\textwidth]{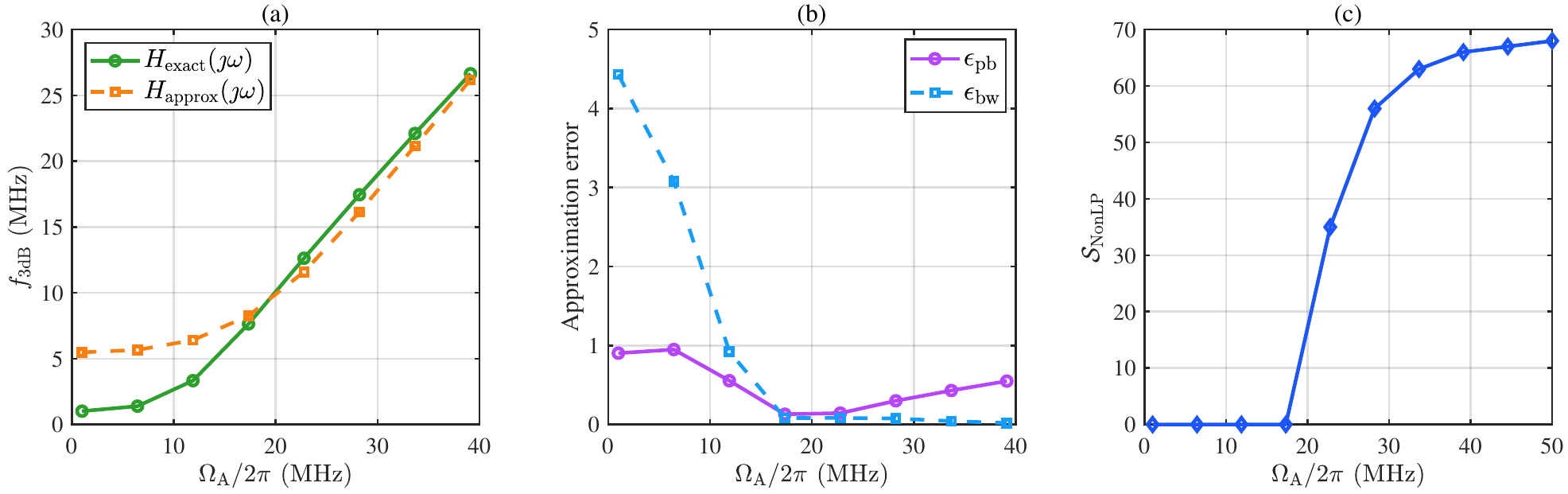}
	\caption{Effect of the auxiliary-field Rabi frequency $\Omega_{\text{A}}$ on 3-dB bandwidth and low-pass regularity. (a) Exact and approximated $f_\text{3dB}$ versus $\Omega_{{\text{A}}}$. (b) Approximation errors $\epsilon_{\text{pb}}$ and $\epsilon_{\text{bw}}$. (c) Non-low-pass score $\mathcal{S}_{\text{NonLP}}$.} \label{f_vs_Omega_A}
\end{figure*}

\begin{figure*}[t]
	\centering
	\includegraphics[width=0.92\textwidth]{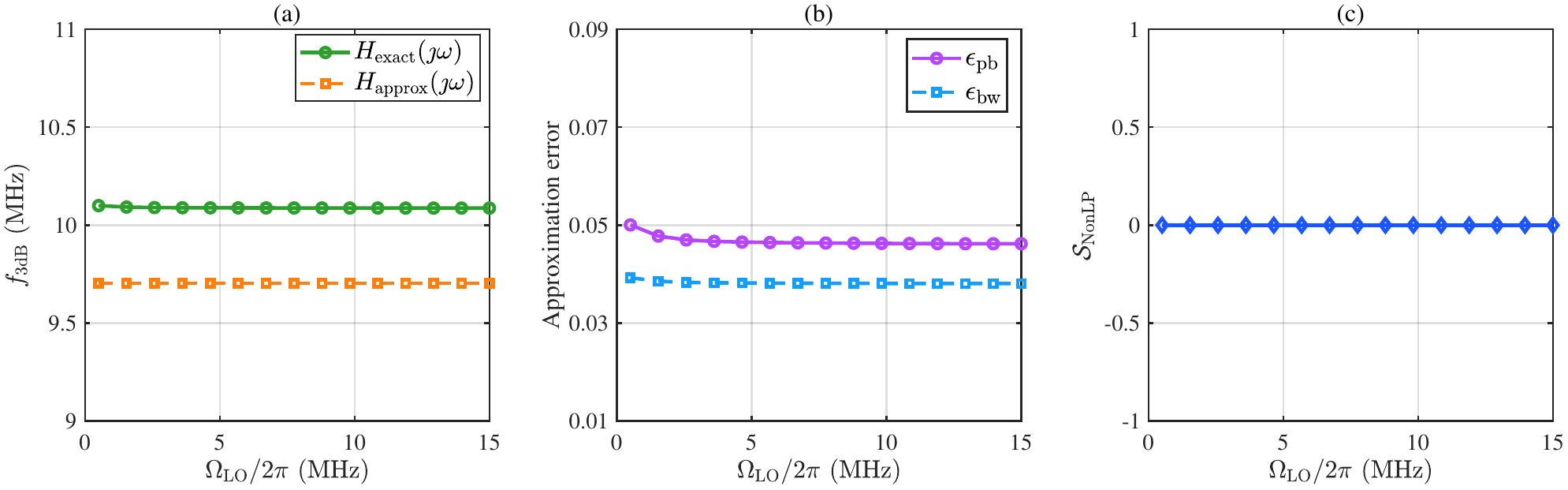}
	\caption{Effect of the LO-field Rabi frequency $\Omega_\text{LO}$ on 3-dB bandwidth and low-pass regularity. 
		(a) Exact and approximated $f_\text{3dB}$ versus $\Omega_{{\text{LO}}}$. 
		(b) Approximation errors  $\epsilon_{\text{pb}}$ and $\epsilon_{\text{bw}}$. 
		(c) Non-low-pass score $\mathcal{S}_{\text{NonLP}}$.} \label{f_vs_Omega_LO}
\end{figure*}

To avoid over-interpreting 3-dB bandwidth in a distorted frequency response, we introduce a low-pass regularity metric, namely non-low-pass score, denoted by $\mathcal{S}_{\text{NonLP}}$. To be specific, let
$\mathcal{H}\left( f \right) = {{\left| {{H_{{\text{exact}}}}\left( {\jmath 2\pi f} \right)} \right|}} / {{\left| {{H_{{\text{exact}}}}\left( 0 \right)} \right|}}$
be the normalized exact detected SWM response sampled over a discrete frequency grid $\{f_i\}_{i=1}^{N_f}$ with $f_{i+1}>f_i$. For an ideal low-pass response, $\mathcal{H}\left( f \right)$ should be monotonically non-increasing. We therefore define
\begin{equation}
{\mathcal{S}_{{\text{NonLP}}}} = \sum\limits_{i = 1}^{{N_f} - 1} \mathbb{I} \left[ {\mathcal{H}\left( {{f_{i + 1}}} \right) - \mathcal{H}\left( {{f_i}} \right) > {\epsilon_{{\text{LP}}}}} \right],
\end{equation}
where $\mathbb{I} \left[  \cdot \right] $ is the indicator function and $\epsilon_{\text{LP}}$ is a small tolerance employed to suppress numerical fluctuations. In this sense, $\mathcal{S}_{\text{NonLP}}=0$ means that the detected SWM response remains monotonic low-pass over the scanned band, whereas a larger $\mathcal{S}_{\text{NonLP}}$ indicates stronger finite-frequency peaking, ripple, or non-monotonic frequency-response distortion.

Fig.~\ref{f_vs_Omega_A} first examines the role of the auxiliary-field Rabi frequency $\Omega_\text{A}$, which directly enters the higher-level dressed kernel of the SWM response. As transpired in Fig.~\ref{f_vs_Omega_A}, increasing $\Omega_\text{A}$ significantly broadens 3-dB bandwidth $f_\text{3dB}$. More specifically, as shown in Fig.~\ref{f_vs_Omega_A}(a), both exact and approximated $f_\text{3dB}$ increase almost monotonically with $\Omega_\text{A}$. However, this bandwidth enhancement is not unlimited. Fig.~\ref{f_vs_Omega_A}(c) portrays that the non-low-pass score $\mathcal{S}_{\text{NonLP}}$ remains zero only within a finite operating window, indicating that the detected SWM response preserves a monotonic low-pass shape in this region. 
Once $\Omega_\text{A}$ is further increased beyond this window, $\mathcal{S}_{\text{NonLP}}$ rises rapidly, which connotes that the response develops finite-frequency peaking or non-monotonic ripple. Therefore, the larger 3-dB bandwidth at very large $\Omega_\text{A}$ should not be interpreted as a clean usable instantaneous bandwidth. This observation explains why the optimized operating point is chosen near the boundary of the strict low-pass region: it provides a bandwidth of approximately 10~MHz while avoiding severe frequency-response distortion.

Fig.~\ref{f_vs_Omega_LO} then investigates the influence of the LO-field Rabi frequency $\Omega_{\text{LO}}$ on the 3-dB bandwidth. Around the SWM operating point, varying $\Omega_{\text{LO}}/2\pi$ from 0.5~MHz to 15~MHz produces only a marginal change in the exact 3-dB bandwidth, which remains close to 10~MHz. The reduced-order two-pole approximation also remains nearly unchanged and consistently predicts a cutoff around 9.7~MHz. Furthermore, both $\epsilon_{\text{bw}}$ and $\epsilon_{\text{pb}}$ stay small, and  $\mathcal{S}_{\text{NonLP}}$ remains zero over the entire scanned range. These results indicate that, once the lower-level transition paths are sufficiently established, $\Omega_{\text{LO}}$ is not the dominant factor for controlling the instantaneous baseband bandwidth.

\begin{figure}[t]
	\centering
	\includegraphics[width=0.49\textwidth]{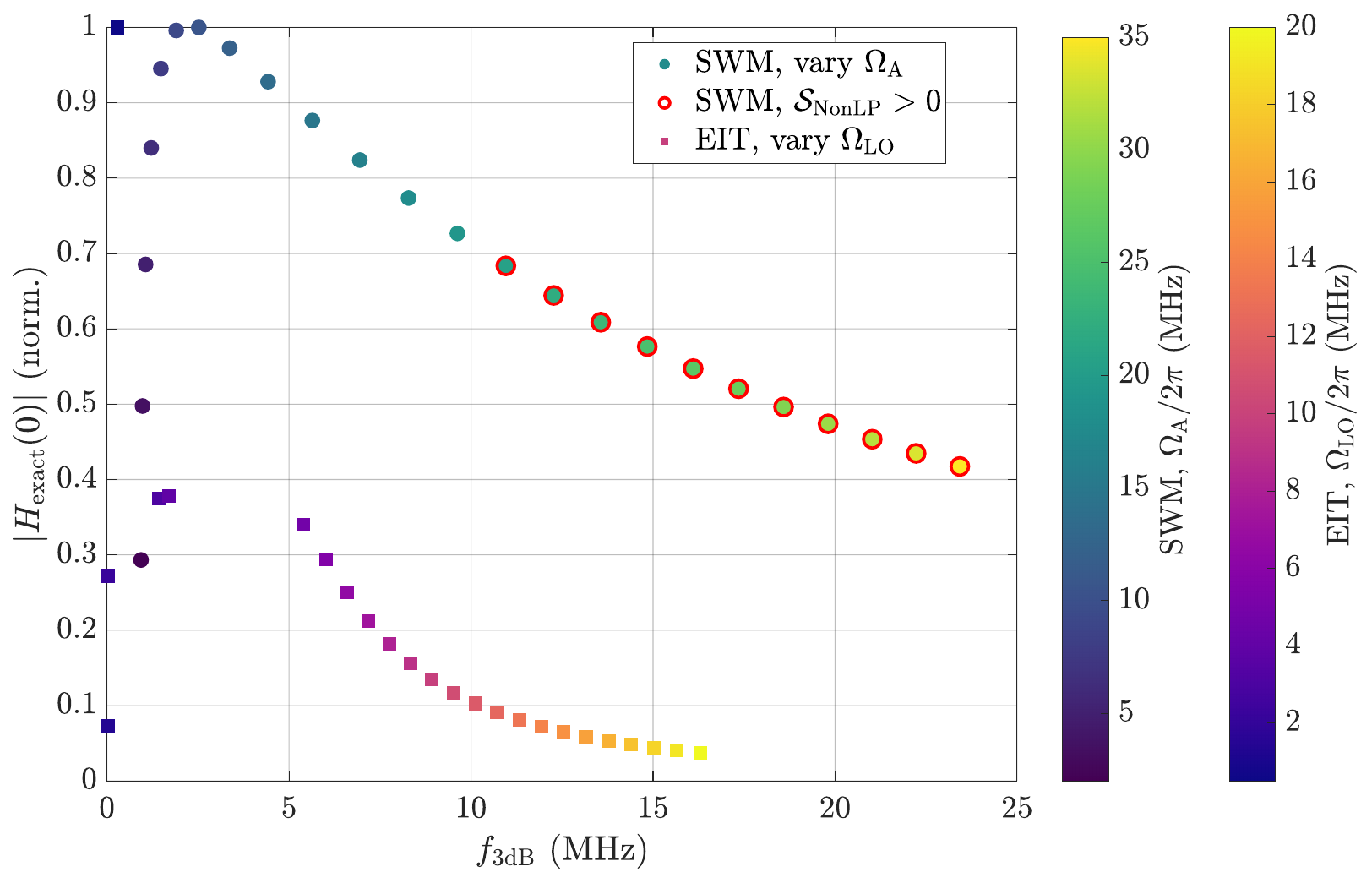}
	\caption{Tradeoff between 3-dB bandwidth and normalized baseband responsivity $\left| H_\text{exact} \left( 0 \right) \right| $ for both EIT and SWM regimes.} \label{BW_responsivity_tradeoff}
\end{figure}

Fig.~\ref{BW_responsivity_tradeoff} compares 3-dB bandwidth $f_\text{3dB}$ and normalized baseband responsivity $\left| H_\text{exact} \left( 0 \right) \right| $ for both EIT-based and SWM-based operating regimes. For the SWM case, frequency $\Omega_{\text{A}}$ is varied, while for the EIT case  $\Omega_{\text{LO}}$ is swept. It can be observed that the EIT-based response is mainly concentrated in the narrow-band region, where increasing $\Omega_{\text{LO}}$ enlarges 3-dB bandwidth $f_\text{3dB}$ but it leads to a rapid reduction in the normalized baseband responsivity. In contrast, the SWM-based response extends to a much wider bandwidth range. As $\Omega_{\text{A}}$ increases, the detected 3-dB bandwidth can be significantly enhanced, while the baseband responsivity gradually decreases, revealing a clear bandwidth-responsivity tradeoff.
Additionally, the red-circled SWM points indicate operating conditions with a nonzero non-low-pass score $\mathcal{S}_\text{NonLP}$, i.e., cases where the response is no longer strictly low-pass-like. These points show that although stronger SWM driving can further push the detected bandwidth, excessive bandwidth enhancement may be accompanied by increased passband distortion. Therefore, the useful wideband operating region should be selected by jointly considering the 3-dB bandwidth, the normalized responsivity, and the low-pass validity of the response.


\subsubsection{Sensitivity}

\begin{figure}[t]
	\centering
	\includegraphics[width=0.48\textwidth]{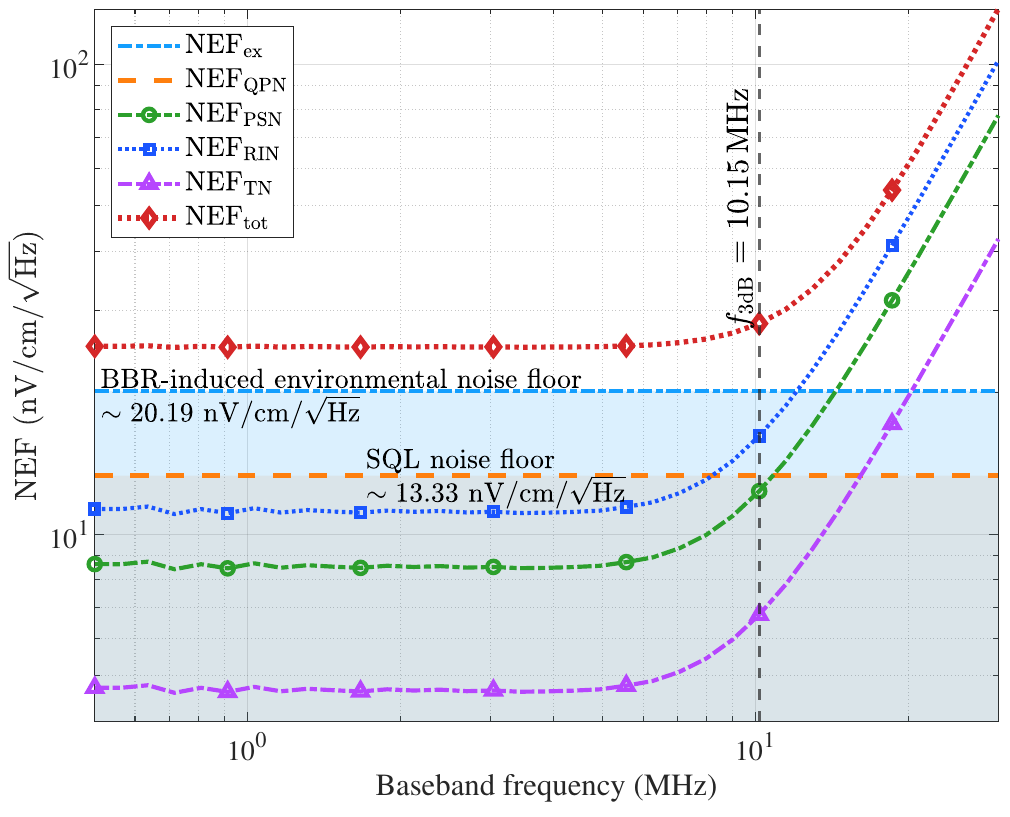}
	\caption{NEF versus baseband frequency.} \label{SWM_NEF}
\end{figure}

To quantify the electric-field sensitivity of the SWM-based Rydberg atomic receiver in a communication-relevant manner, we adopt the NEF framework presented in Sec.~\ref{sec_sensitivity_analysis}. Fig.~\ref{SWM_NEF} plots the NEF as a function of the baseband frequency. For each frequency, the small-signal RF-to-optical transduction gain is extracted from the demodulated optical coherence, and each noise contribution is then referred back to the incident RF field. The BBR-induced environmental noise floor and the SQL-related internal noise floor are also shown as reference levels, which are approximately $20.19~\text{nV/cm}/\sqrt{\text{Hz}}$ and $13.33~\text{nV/cm}/\sqrt{\text{Hz}}$, respectively.
At low baseband frequencies, all NEF components remain nearly flat, and the total NEF is maintained at the level of a few tens of $\text{nV/cm}/\sqrt{\text{Hz}}$. In this region, the RF-to-optical conversion gain is sufficiently high, so technical readout noises (PSN/RIN/TN) are effectively suppressed and the residual limitation is primarily due to fundamental atomic/environmental noise. This trend is not only due to an intrinsic increase of these noise sources themselves, but rather attributes to the $\text{NEF}_\text{tot}$ with frequency. More specifically, when the effective RF-to-optical conversion magnitude decreases, the same detection-chain noise corresponds to a larger input-referred field. Furthermore, the 3-dB bandwidth is $f_\text{3dB} = 10.15~\text{MHz}$, beyond which the sensitivity penalty becomes increasingly severe.

\begin{figure}[t]
	\centering
	\includegraphics[width=0.49\textwidth]{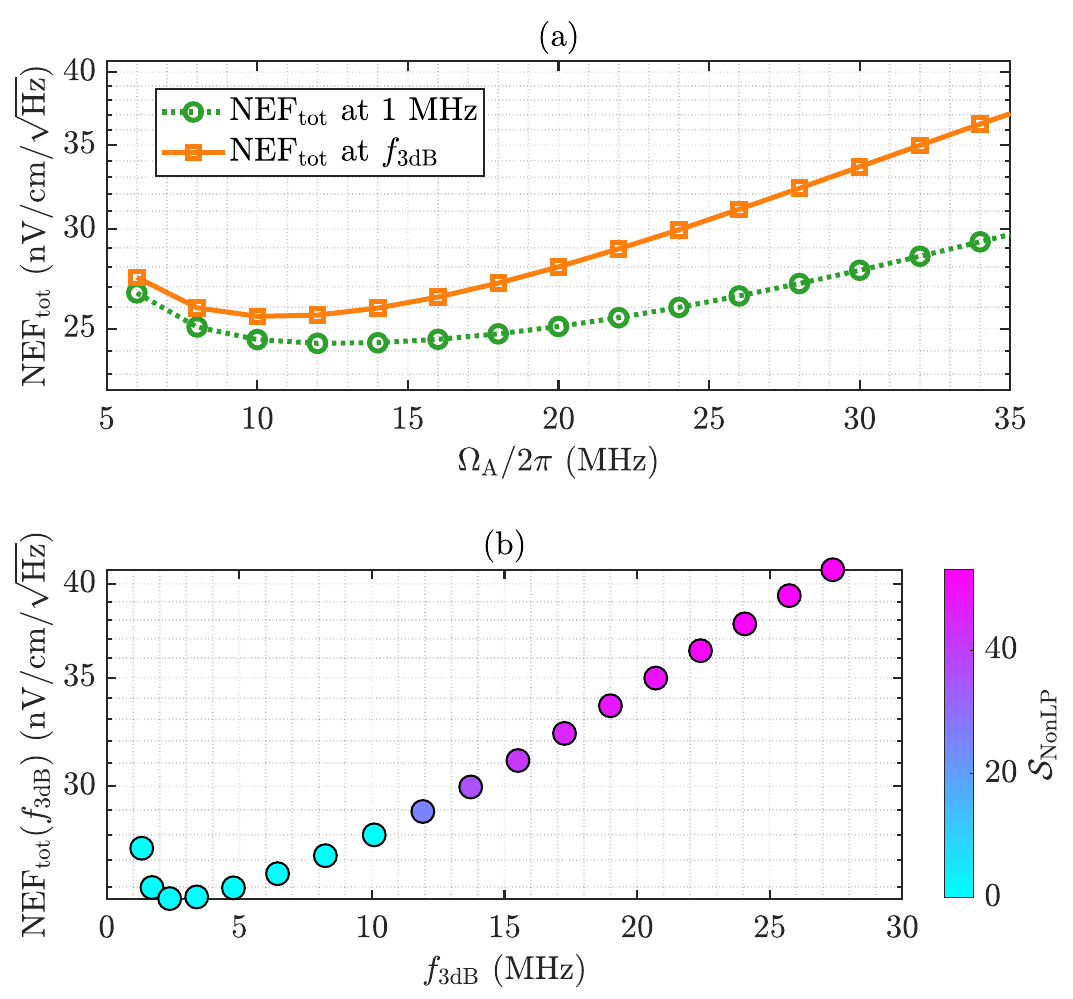}
	\caption{Sensitivity-bandwidth tradeoff. 
		(a) $\text{NEF}_\text{tot}$ at a fixed low baseband frequency of $1~\text{MHz}$ and at 3-dB bandwidth $f_{3\text{dB}}$ as functions of $\Omega_\text{A}$. 
		(b) Tradeoff between 3-dB bandwidth $f_\text{3dB}$, sensitivity $\text{NEF}_\text{tot}$, and the non-low-pass score $\mathcal{S}_{\text{NonLP}}$. 
	}\label{NEF_bandwidth_OmegaA_tradeoff}
\end{figure}

Fig.~\ref{NEF_bandwidth_OmegaA_tradeoff} further portrays the sensitivity-bandwidth tradeoff induced by the auxiliary-field Rabi frequency $\Omega_\text{A}$. More specifically, Fig.~\ref{NEF_bandwidth_OmegaA_tradeoff}(a) delivers two representative sensitivity curves, i.e., $\text{NEF}_{\text{tot}}$ at a fixed low baseband frequency of $1~\text{MHz}$ and $\text{NEF}_{\text{tot}}$ evaluated at $f_{3\text{dB}}$. At small-to-moderate $\Omega_\text{A}$, both curves first slightly decrease and then remain at the level of several tens of $\text{nV/cm}/\sqrt{\text{Hz}}$, indicating that a moderate $\Omega_{{\text{A}}}$ can improve the RF-to-optical conversion response without introducing a severe sensitivity penalty. However, as $\Omega_\text{A}$ is further increased, both curves gradually degrade. 
Fig.~\ref{NEF_bandwidth_OmegaA_tradeoff}(b) gives a more compact view of the coupled bandwidth-sensitivity-regularity tradeoff. As it transpired in Fig.~\ref{NEF_bandwidth_OmegaA_tradeoff}(b), when $f_{3\text{dB}}$ is increased from the narrowband regime to the 10-MHz-class regime, the $\text{NEF}_{\text{tot}}$ remains moderate and $\mathcal{S}_{\text{NonLP}}$ stays close to zero, implying that the bandwidth enhancement is still associated with a clean monotonic low-pass response. Beyond this regime, further increasing $\Omega_\text{A}$ continues to increase the apparent first-crossing 3-dB bandwidth, but $\text{NEF}_{\text{tot}} \left(f_{3\text{dB}}\right)$ rises rapidly and $\mathcal{S}_{\text{NonLP}}$ becomes large. Therefore, the largest bandwidth does not correspond to the best operating point. Instead, the practical SWM operating point should be selected by jointly considering bandwidth, sensitivity, and low-pass regularity.

\subsection{Linear Dynamic Range}
\subsubsection{P1dB and IIP3 Performance}

\begin{figure}[t]
	\centering
	\includegraphics[width=0.49\textwidth]{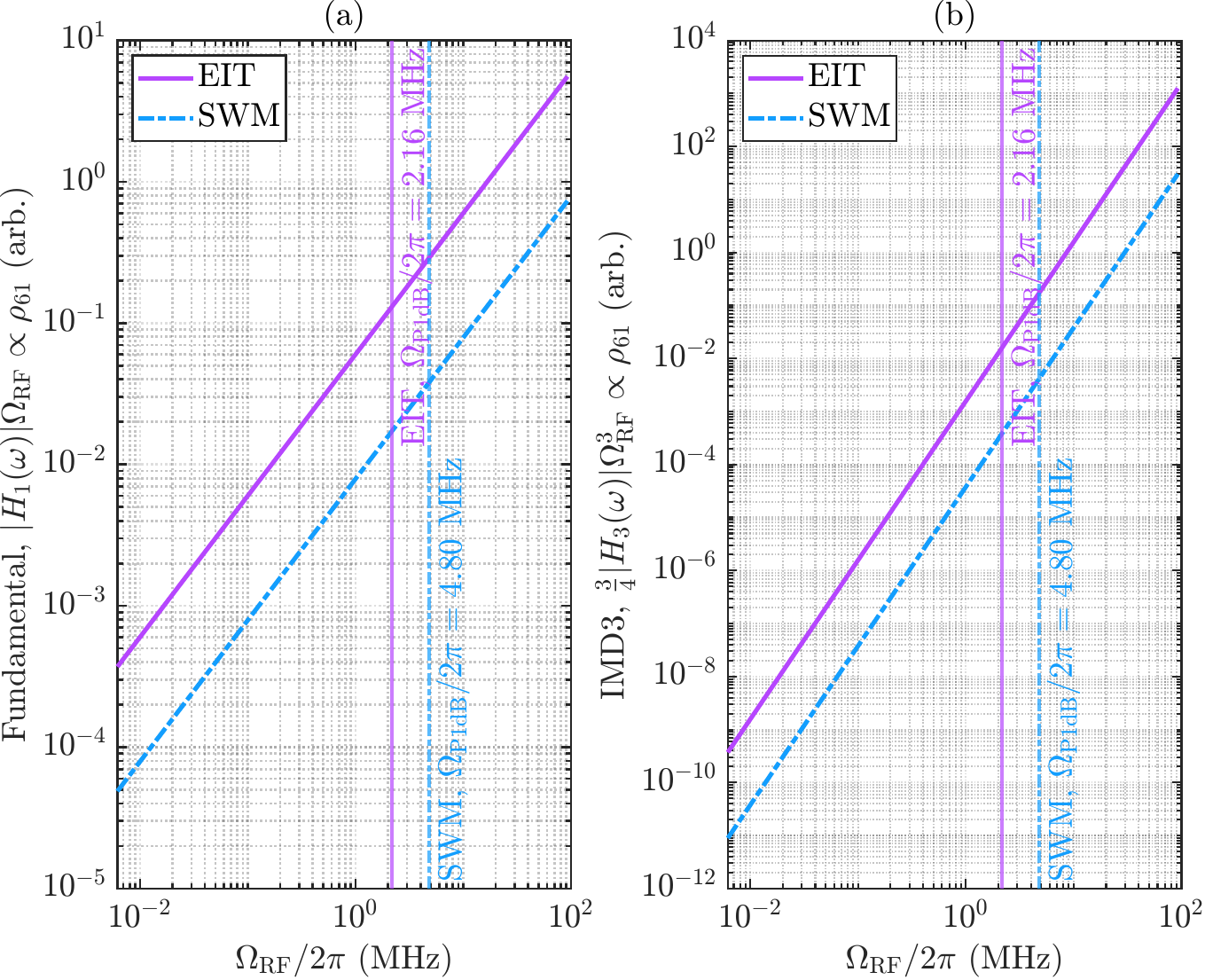}
	\caption{(a) Fundamental components versus the RF Rabi frequency $\Omega_{\text{RF}}$. (b) IMD3 components versus $\Omega_{\text{RF}}$.} \label{fundamental_IMD3_components}
\end{figure}

\begin{figure}[t]
	\centering
	\includegraphics[width=0.49\textwidth]{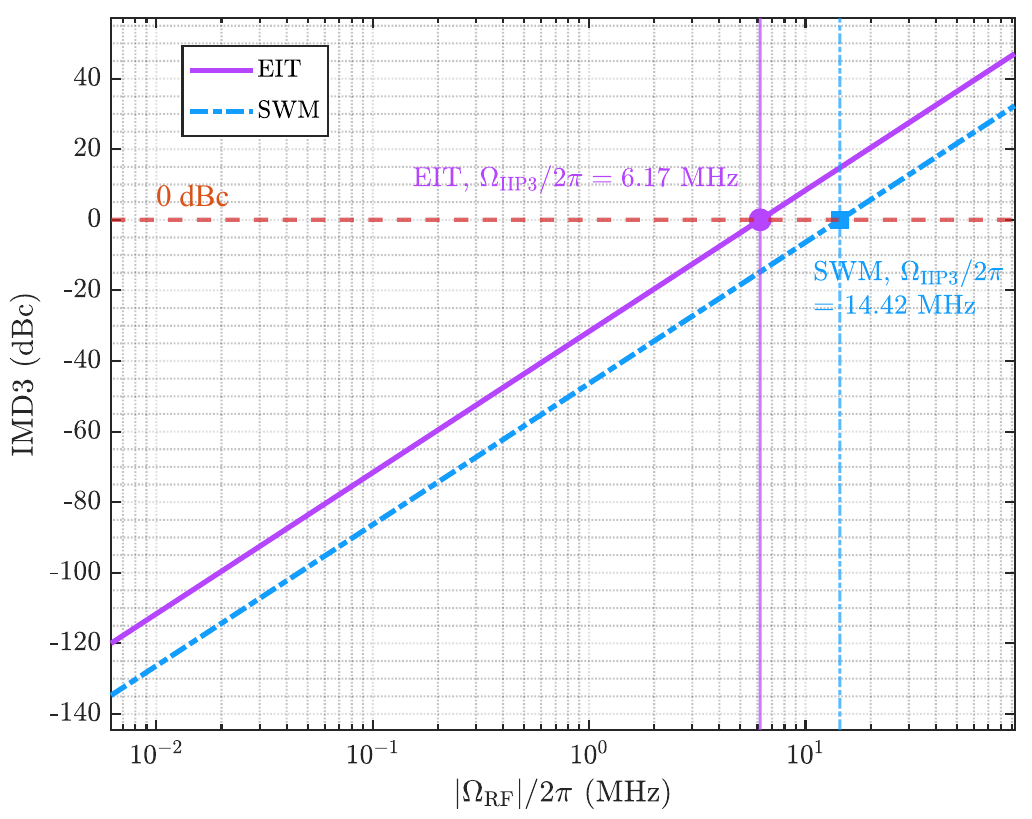}
	\caption{IMD3 performance in dBc versus the RF Rabi frequency $\Omega_{\text{RF}}$.} \label{Fig3_IMD3_IIP3}
\end{figure}

Fig.~\ref{fundamental_IMD3_components} and Fig.~\ref{Fig3_IMD3_IIP3} jointly quantify the third-order nonlinearity of the EIT- and SWM-based Rydberg atomic receivers, while bridging the quantum model to standard RF linearity metric. Taken together, they aim to address two key questions: (i) whether the atomic response behaves as a well-defined third-order nonlinearity over the input range of interest; and (ii) for a given RF driven level, which architecture delivers better multi-tone linearity in terms of conventional IMD3/IIP3 measures.

Fig.~\ref{fundamental_IMD3_components}(a) and Fig.~\ref{fundamental_IMD3_components}(b) illustrate the fundamental component  ${H_1}\left( \omega  \right){\Omega _{{\text{RF}}}}$ and the IMD3 component 
$\frac{3}{4}{H_3}\left( \omega  \right)\Omega _{{\text{RF}}}^3$, respectively, versus the RF Rabi frequency $\Omega_{{\text{RF}}}$. The vertical lines in both subfigures mark the P1dB. As can be observed in Fig.~\ref{fundamental_IMD3_components}(a), the EIT baseline exhibits a larger fundamental response than the SWM configuration over the scanned input range. This indicates that, within the operating region, the EIT-based Rydberg atomic receiver provides a stronger narrowband small-signal response. However, the fundamental amplitude alone cannot be used to evaluate the input linear range, because the nonlinear distortion must be evaluated relative to the desired fundamental component.
Fig.~\ref{fundamental_IMD3_components}(b) shows the corresponding IMD3 components. The EIT configuration also exhibits a larger absolute IMD3 amplitude than SWM, implying that the larger EIT fundamental response is accompanied by stronger third-order nonlinear effect. By contrast, although the SWM response has a smaller absolute fundamental amplitude, its IMD3 component is also substantially lower. This indicates that the selected SWM wideband operating point suppresses the third-order distortion more effectively.
Furthermore, for a given $f_\text{3dB}$, the P1dB achieved by the SWM scheme lies at a higher $\Omega_{\text{RF}}$ than that of the EIT counterpart. This indicates that the SWM configuration provides a wider  linear range.

Fig.~\ref{Fig3_IMD3_IIP3} further presents the same third-order distortion in the RF engineer's language. It plots IMD3 in dBc, defined in~(\ref{IMD3}), versus $\Omega_{{\text{RF}}}$ on a logarithmic scale. Extrapolating each line to $0~\text{dBc}$ gives the corresponding IIP3, i.e., ${\Omega _{{\text{IIP3}}}}/2\pi  = 14.42~{\text{MHz}}$ for SWM and ${\Omega _{{\text{IIP3}}}}/2\pi  = 6.17~{\text{MHz}}$ for EIT. 
The SWM scheme exhibits a higher IIP3 value and thus the wider dynamic linear range. This substantiates that the optimized SWM wideband operating point can maintain a favorable linear range while providing a substantially larger detected 3-dB bandwidth. Therefore, the linearity comparison should be interpreted jointly with bandwidth and responsivity, rather than as an isolated metric, reflecting an overall bandwidth-responsivity-linearity tradeoff.


\subsubsection{Trade-off Between Bandwidth and IIP3}

\begin{figure}[t]
	\centering
	\includegraphics[width=0.49\textwidth]{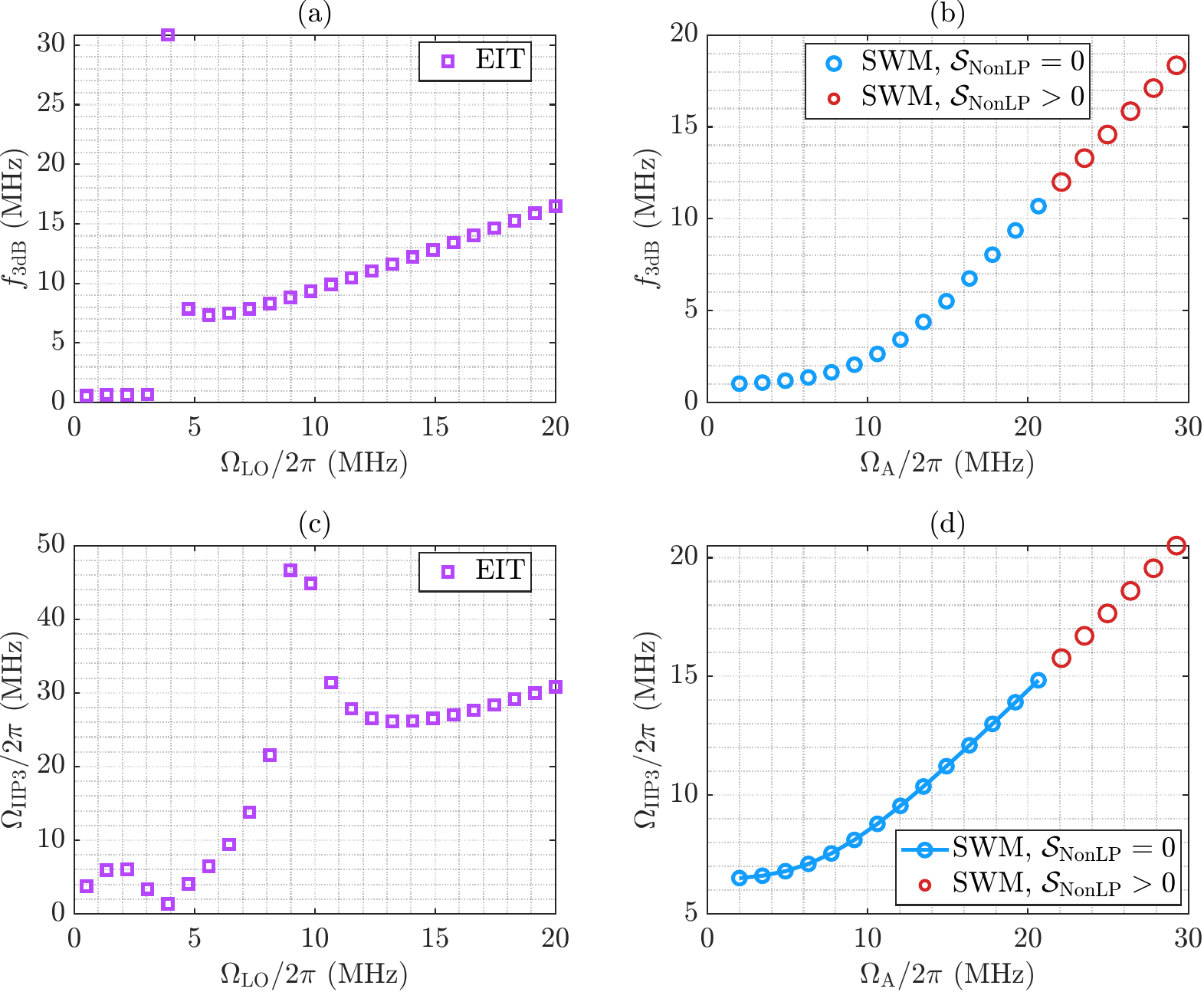}
	\caption{Dependence of 3-dB bandwidth $f_\text{3dB}$ and IIP3 on the Rabi frequencies $\Omega_{\text{LO}}$ and $\Omega_{\text{A}}$ for EIT and SWM configurations. (a) EIT bandwidth versus $\Omega_{\text{LO}}$. 
		(b) SWM bandwidth versus $\Omega_{\text{A}}$. 
		(c) EIT IIP3 versus $\Omega_{\text{LO}}$. 
		(d) SWM IIP3 versus $\Omega_{\text{A}}$. } \label{Bandwidth_IIP3}
\end{figure}

Fig.~\ref{Bandwidth_IIP3} examines how the RF signal bandwidth and third-order linearity evolve when we tune the key control parameters in a pair of configurations: the LO Rabi frequency $\Omega_{\text{LO}}$ in the EIT scheme and the auxiliary optical Rabi frequency $\Omega_{\text{A}}$ in the SWM scheme. Our goal is to identify which knob can genuinely expand the baseband (envelope) bandwidth, and to what extent this can be achieved without sacrificing linearity.

As shown in Fig.~\ref{Bandwidth_IIP3}(a), the EIT bandwidth is highly sensitive to $\Omega_{\text{LO}}$. At small $\Omega_{\text{LO}}$, the detected SWM response remains narrowband. As $\Omega_{\text{LO}}$ increases, $f_\text{3dB}$ increases, but the response may also enter a non-low-pass regime due to dressed-state splitting and finite-frequency peaking. In contrast, Fig.~\ref{Bandwidth_IIP3}(b) shows that the SWM bandwidth increases more smoothly with $\Omega_{\text{A}}$. In the strict low-pass region, marked by $\mathcal{S}_{\text{NonLP}}=0$, increasing $\Omega_{\mathrm{A}}$ enlarges the bandwidth from the MHz level to the 10-MHz-class regime. This confirms that $\Omega_{\text{A}}$ provides an effective tuning knob for broadening the SWM baseband response. However, when $\Omega_{\text{A}}$ becomes too large, the response enters a non-low-pass regime, in which larger bandwidth can be spotted, but they are not preferred operating points because the response is no longer a clean low-pass transfer function.

Both Figs.~\ref{Bandwidth_IIP3}(c) and~\ref{Bandwidth_IIP3}(d) demonstrate IIP3 trends. For the EIT baseline, $\Omega_{\text{IIP3}}$ varies non-monotonically with $\Omega_{\text{LO}}$. Certain LO settings yield a high IIP3, but these points are associated with operating regimes where the bandwidth behavior is also strongly affected by non-low-pass peaking. This indicates that improving EIT linearity by increasing the dressing strength does not naturally lead to a wide and clean baseband response. For the SWM configuration shown in Fig.~\ref{Bandwidth_IIP3}(d), the IIP3 increases gradually with $\Omega_{\text{A}}$ in the strict low-pass region, indicating that the wideband operating condition does not necessarily sacrifice the linear range. In particular, the points with $\mathcal{S}_{\text{NonLP}}=0$ show a simultaneous increase in both $f_{3\text{dB}}$ and $\Omega_{\text{IIP3}}$ before the onset of non-low-pass distortion. This behavior suggests that a properly chosen auxiliary field can improve both the detected bandwidth and the input-referred linearity.


\begin{figure}[t]
	\centering
	\includegraphics[width=0.49\textwidth]{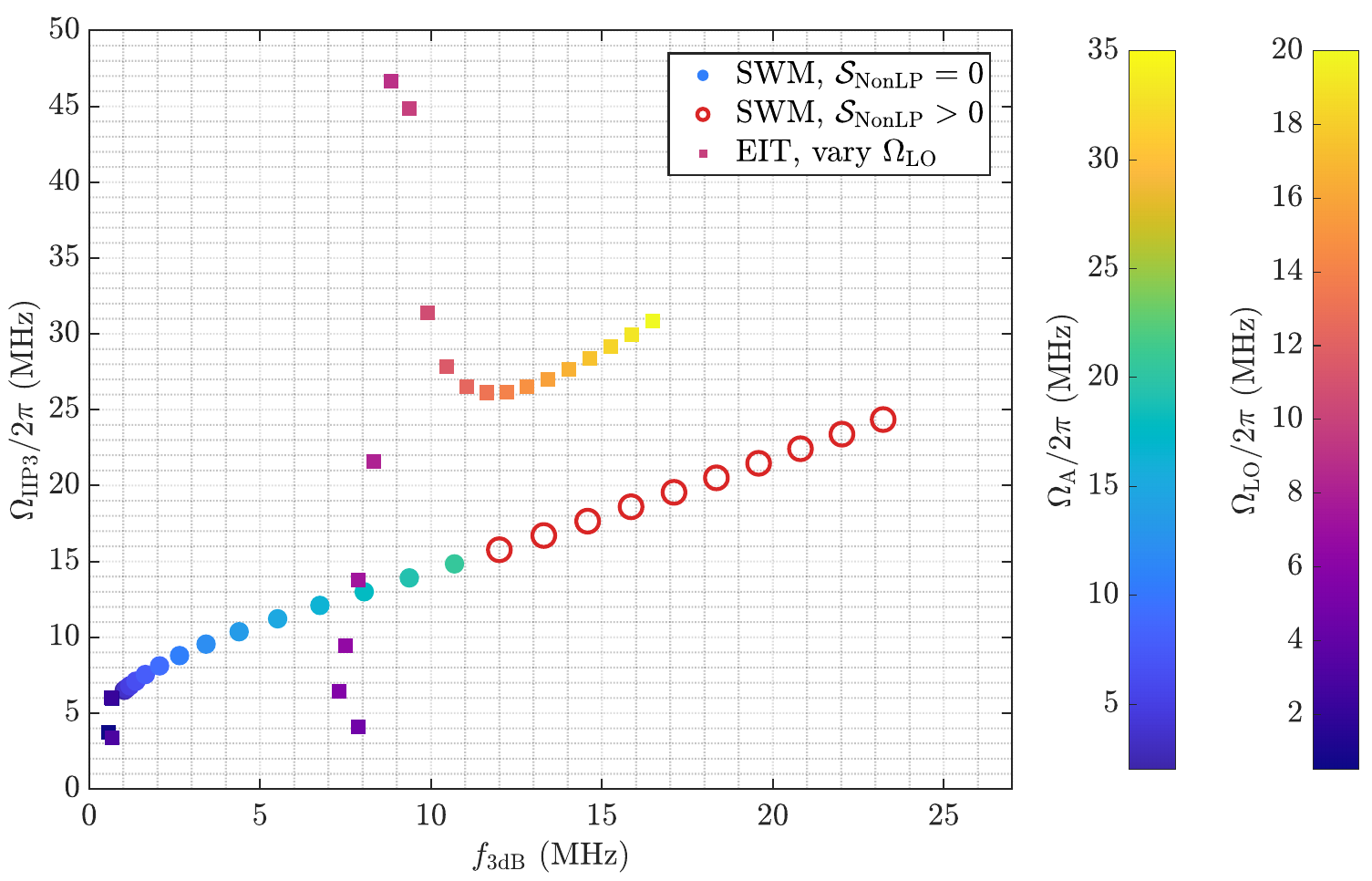}
	\caption{Trade-off between EIT and SWM schemes with respect to 3-dB bandwidth $f_\text{3dB}$ and IIP3 $\Omega_\text{IIP3}$.} \label{tradeoff_IIP3}
\end{figure}

Fig.~\ref{tradeoff_IIP3} intends to compare, on the same footing, how much linearity (evaluated by IIP3, $\Omega_\text{IIP3}$) can be achieved for a given effective 3-dB bandwidth $f_\text{3dB}$ when the Rydberg atomic receiver is operated either in a conventional super-heterodyne EIT technique (squares) or in the proposed SWM mode (circles). In other words, it visualizes the design space available to a system engineer: for a target bandwidth, how far can we push the input dynamic range by tuning the driven light field $\Omega_{\text{LO}}$ or $\Omega_{\text{A}}$. 

Specifically, the SWM points form an almost smooth, rising trend: as $\Omega_{\text{A}}$ increases, both $f_\text{3dB}$ and $\Omega_\text{IIP3}$ increase in a fairly regular way. When $\Omega_{\text{A}}$ becomes excessively large, the response enters a non-low-pass regime. Although these points may exhibit even larger bandwidth and IIP3, they are not treated as preferred operating points because the baseband response is no longer a clean monotonic low-pass response.
This indicates a benign bandwidth–linearity relationship for SWM, where moving to a wider bandwidth may not require sacrificing IIP3 within the operating range. Physically, $\Omega_{\text{A}}$ in the SWM scheme mainly controls the overall strength and power broadening of the SWM pathway: a stronger $\Omega_{\text{A}}$ enhances the useful third-order susceptibility responsible for detection, while higher-order distortion terms grow more slowly, so the ratio between fundamental and inter-modulation components improves.

By contrast, the EIT data exhibits a more structured pattern. For weak $\Omega_{\text{LO}}$, the system remains narrowband and only moderately linear. As $\Omega_{\text{LO}}$ is increased to approximately $6~\text{MHz}$, both bandwidth and IIP3 improve; however, beyond this region, the IIP3 shows a pronounced peak followed by a saturation. This can be attributed to interference among multiple RF-induced paths contributing to the third-order response. More explicitly, at a specific $\Omega_{\text{LO}}$, the third-order distortion components generated along different pathways nearly cancel each other, resulting in a strong suppression of inter-modulation products while the desired linear susceptibility remains large. A slight detuning from this operating point disrupts this cancellation, causing the IIP3 to drop rapidly.

\subsection{Discussion}

The above simulation results suggest that EIT-based and SWM-based Rydberg atomic receivers should be treated as two complementary operating regimes rather than as one scheme universally outperforming the other. More specifically, the EIT configuration is naturally applicable to narrowband sensing and reception. As validated in Fig.~\ref{BW_responsivity_tradeoff}, EIT can provide a relatively strong baseband responsivity in the low-bandwidth regime, which is beneficial for narrowband sensing applications. In addition, Figs.~\ref{Bandwidth_IIP3}(c) and~\ref{tradeoff_IIP3} show that certain EIT operating points can achieve higher IIP3 values. However, these points are strongly dependent on $\Omega_{\text{LO}}$ and are often associated with narrowband or non-low-pass response behavior. Therefore, EIT is preferable when the signal bandwidth is modest and the receiver can be carefully biased at a narrowband operating point.

By contrast, the SWM configuration is more suitable for wideband RF-to-optical transduction. Fig.~\ref{3dB_illustration} conveys that the selected SWM operating point can achieve 3-dB bandwidth of approximately 10~MHz, which is about one order of magnitude larger than the EIT benchmark. Fig.~\ref{f_vs_Omega_A} further confirms that the auxiliary-field Rabi frequency $\Omega_{\text{A}}$ acts as an effective bandwidth-control knob. Within the strict low-pass region, i.e., $\mathcal{S}_{\text{NonLP}}=0$, increasing $\Omega_{\text{A}}$ broadens the detected SWM response without inducing strong distortion. Additionally, Figs.~\ref{fundamental_IMD3_components}-\ref{tradeoff_IIP3} show that the SWM configuration can maintain a favorable linear range, as reflected by its larger P1dB and IIP3 at the selected wideband operating point. Nevertheless, this is not to say that the SWM bandwidth enhancement is cost-free. Fig.~\ref{BW_responsivity_tradeoff} indicates that increasing SWM-associated bandwidth reduces the normalized baseband responsivity, while Fig.~\ref{NEF_bandwidth_OmegaA_tradeoff} presents that excessively large $\Omega_{\text{A}}$ can degrade NEF and drive the response into a non-low-pass regime. Therefore, the best SWM operating point is not the one with the largest bandwidth, but the one that jointly provides large $f_{3\text{dB}}$, acceptable NEF, high IIP3, and $\mathcal{S}_{\text{NonLP}}=0$.

From an engineering perspective, EIT is a more responsive choice for narrowband applications, whereas SWM is more attractive for wideband reception. The key benefit of SWM is that it expands the feasible design space: using the proposed closed-form two-pole low-pass framework, SWM enables wideband operation while retaining high sensitivity and favorable linearity, provided that the auxiliary-field strength is selected under the low-pass regularity constraint.

\section{Conclusion}\label{Sec_V_conclusion}

In this work, we have developed a wideband quantum-transduction framework for a SWM-based Rydberg atomic receiver and established a baseband response model that bridges atomic coherence dynamics to communication-level performance metrics. Specifically, under the near-resonant weak-signal regime of interest, the exact detected SWM response is reduced to a closed-form two-pole low-pass approximation. This reduced-order model provides an analytically tractable characterization of the detected 3-dB bandwidth and reveals how the bandwidth is manipulated by the dressed higher-level atomic dynamics and the relevant optical/RF parameters.
To justify the use of this approximation, we further examined its validity range through approximation-error and low-pass-regularity analyses. We evince that the proposed reduced-order approximation accurately captures both the detected bandwidth and the passband shape within properly selected operating regions. In contrast, excessive detuning or overly strong dressing may induce undesirable non-low-pass features and consequently degrade the approximation accuracy.

Additionally, simulation results demonstrate that SWM can achieve a 3-dB bandwidth of approximately 10~MHz while maintaining favorable linearity and sensitivity under the strict low-pass condition. The comparison with the EIT baseline indicates that the two schemes should be treated as complementary rather than universally ordered. EIT remains attractive for narrowband sensing and reception due to its stronger local responsivity, whereas SWM expands the feasible design space for wideband RF-to-optical transduction by jointly tuning bandwidth, sensitivity, and linearity through the auxiliary-field control. From an engineering perspective, the preferred SWM operating region is therefore not the one with the largest available bandwidth, but the one that simultaneously provides a large bandwidth, acceptable NEF, favorable IIP3, and low-pass regularity.

Future work will focus on experimentally validating the proposed closed-form two-pole design framework using SWM-based Rydberg receiver prototypes. This includes calibrating the noise model, verifying the predicted bandwidth-linearity trade-off, and evaluating the receiver under realistic wideband communication waveforms.

\appendices
\section{Derivation of ${\rho _{61}}\left( \omega  \right)$ in (\ref{rho_61})}\label{appendix_rho_61}

We derive here the steady-state solution for $\rho_{61}$ that underlies the SWM process. Starting from the master equation in~(\ref{master_equation}) and working in the weak-probe and weak-RF regime, we keep only the slowly varying coherences $\rho_{j1}$ with $j = 2,...,6$. Their equations can be written as
\begin{subequations}
\begin{align}
{{\dot \rho }_{21}} &=  - {D_2}{\rho _{21}} + \frac{\jmath }{2}{\Omega _{\text{P}}}, \\
{{\dot \rho }_{31}} &=  - {D_3}{\rho _{31}} + \frac{\jmath }{2}{\Omega _{\text{C}}}{\rho _{21}}, \\
{{\dot \rho }_{41}} &=  - {D_4}{\rho _{41}} + \frac{\jmath }{2}{\Omega _{{\text{LO}}}}{\rho _{31}},\\
{{\dot \rho }_{51}} &=  - {D_5}{\rho _{51}} + \frac{\jmath }{2}{\Omega _{{\text{RF}}}}\left( t \right){\rho _{41}} + \frac{\jmath }{2}{\Omega _{\text{A}}}\left( t \right){\rho _{61}},\\
{{\dot \rho }_{61}} &=  - {D_6}{\rho _{61}} + \frac{\jmath }{2}\Omega _{\text{A}}^*{\rho _{51}}.
\end{align}
\end{subequations}
By taking the Fourier transform of the above equations, we obtain algebraic relations for the first-order response at analysis frequency $\omega$, yielding that 
\begin{subequations}
	\begin{align}
{D_2}\left( \omega  \right){\rho _{21}}\left( \omega  \right) &= \frac{\jmath }{2}{\Omega _{\text{P}}}, \label{Fourier_1}\\
{D_3}\left( \omega  \right){\rho _{31}}\left( \omega  \right) &= \frac{\jmath }{2}{\Omega _{\text{C}}}{\rho _{21}}\left( \omega  \right),\label{Fourier_2}\\
{D_4}\left( \omega  \right){\rho _{41}}\left( \omega  \right) &= \frac{\jmath }{2}{\Omega _{{\text{LO}}}}{\rho _{31}}\left( \omega  \right),\label{Fourier_3}\\
{D_5}\left( \omega  \right){\rho _{51}}\left( \omega  \right) &= \frac{\jmath }{2}{\Omega _{{\text{RF}}}}\left( \omega  \right){\rho _{41}}\left( \omega  \right) + \frac{\jmath }{2}{\Omega _{\text{A}}}{\rho _{61}}\left( \omega  \right),\label{Fourier_4}\\
{D_6}\left( \omega  \right){\rho _{61}}\left( \omega  \right) &= \frac{\jmath }{2}\Omega _{\text{A}}^*{\rho _{51}}\left( \omega  \right) \label{Fourier_5} .
	\end{align}
\end{subequations}
Since the relevant analysis frequencies $\omega$ are much smaller than the optical detunings and dephasing rates of the first three levels, we may, to an excellent approximation, neglect the $\omega$-dependence in $D_2$, $D_3$, $D_4$ and treat $\rho_{41} $ as frequency independent. We keep the explicit $\omega$ dependence only in $D_5 \left( \omega \right)$ and $D_6 \left( \omega \right)$, which are responsible for the bandwidth of interest. 

Equations~(\ref{Fourier_1}), (\ref{Fourier_2}), and (\ref{Fourier_3}) can be solved recursively, and we obtain ${\rho _{41}}$ as follows
\begin{equation}\label{tmp_rho_41}
{\rho _{41}} = {\left( {\frac{\jmath }{2}} \right)^3}\frac{{{\Omega _{\text{P}}}{\Omega _{\text{C}}}{\Omega _{{\text{LO}}}}}}{{{D_2}{D_3}{D_4}}}.
\end{equation}
Both equations in (\ref{Fourier_4}) and (\ref{Fourier_5}) yields
\begin{equation}\label{tmp_rho_51_41}
\left[ {{D_5}\left( \omega  \right) + \frac{{{{\left| {{\Omega _A}} \right|}^2}}}{{4{D_6}\left( \omega  \right)}}} \right]{\rho _{51}}\left( \omega  \right) = \frac{\jmath }{2}{\Omega _{{\text{RF}}}}\left( \omega  \right){\rho _{41}}\left( \omega  \right).
\end{equation}
By substituting (\ref{tmp_rho_41}) into (\ref{tmp_rho_51_41}), we finally obtain the closed-form expression presented in~(\ref{rho_61}).

\section{Error Analysis of the Lower-Level Factor Approximation}\label{appendix_error_analysis}

We define the normalized factor correction term
\begin{equation}
	\mathfrak{P} \left( \omega \right) \triangleq
	\prod_{j=2}^{4}
	\left(
	1+\frac{\jmath\omega}{D_j^{{\left( 0 \right)}}}
	\right)^{-1}.
\end{equation}
Then the exact lower-level factor can be written as
\begin{equation}
	\frac{1}{ D_2\left( \omega \right) D_3 \left( \omega\right) D_4 \left(\omega \right)}
	=
\frac{\mathfrak{P}\left( \omega \right)}{  {D_2^{{\left( 0 \right)}}D_3^{{\left( 0 \right)}}D_4^{{\left( 0 \right)}}}},
\end{equation}
whereas the reduced-order approximation corresponds to replacing
$\mathfrak{P}\left( \omega \right) \approx 1$.
To quantify the approximation error, we define the absolute factor error as
$\varepsilon_{\mathfrak{P}}\left( \omega \right)\triangleq \left| \mathfrak{P} \left( \omega \right)-1 \right|$.
Using the first-order expansion $\left(1+x\right)^{-1}=1-x+\mathcal{O}\left( x^2\right) $, we obtain, for sufficiently small $\omega$,
\begin{align}
	\mathfrak{P} \left( \omega \right) =
	\prod_{j=2}^{4}
	\left(
	1+\frac{\jmath\omega}{D_j^{{\left( 0 \right)}}}
	\right)^{-1} 
	\approx
	1-\jmath\omega\sum_{j=2}^{4}\frac{1}{D_j^{{\left( 0 \right)}}}+\mathcal{O} \left( \omega^2 \right) .
\end{align}
Hence, it yields that
\begin{equation}
	\mathfrak{P} \left( \omega \right) - 1
	\approx
	-\jmath \omega \sum_{j=2}^{4}\frac{1}{D_j^{{\left( 0 \right)}}}+\mathcal{O}\left( \omega^2\right) ,
\end{equation}
which yields the bound
\begin{equation}
	\varepsilon_{\mathfrak{P}}\left( \omega \right)	= \left| \mathfrak{P} \left( \omega \right)-1 \right|
	\lesssim
	\left| \omega \right| \sum_{j=2}^{4}\frac{1}{\left| D_j^{{\left( 0 \right)}}\right| }
	+\mathcal{O}\left( \omega^2\right) .
\end{equation}
Therefore, a sufficient condition for the lower-level factor to remain approximately frequency-flat over the passband of interest is
\begin{equation}
	\left| \omega \right| \sum_{j=2}^{4}\frac{1}{|D_j^{(0)}|}\ll 1.
\end{equation}
Under this condition, the $\omega$-dependence contributed by $D_2\left( \omega\right) $, $D_3\left( \omega\right) $, and $D_4\left( \omega\right) $ only appears as a higher-order correction, and the dominant roll-off of the reduced-order response is determined by the final-stage dressed kernel.

\bibliographystyle{IEEEtran}
\bibliography{Ref/ref_Rydberg_SWM}

\end{document}